# Ultrafast Response of Plasmonic Nanostructures

Sunil Kumar[*,#] and A. K. Sood

## 1 Introduction

Plasmonic metal nanostructures are an integral part of nanophotonic device applications owing to their ability to generate strong localized electromagnetic fields when illuminated from the far-field. These nanostructures can be prepared in the laboratory with precise shape and size by various chemical and physical synthesis techniques as demonstrated in recent years. Typically, the size of these structures is in the range of few nm to a few hundred nm, i.e., much below the wavelength of light of observation or excitation. The–focus of this chapter is to exploit the ultrafast response of metal nanostructures following femtosecond laser pulse excitation. Femtosecond laser pulses are very important experimental probes to explore the material response at an ultrafast time scale (~10-10,000 fs) on the order of the electronic and vibrational wave packet dynamics in solids and molecules. Broadly, after ultrafast optical excitation the system of study undergoes various microscopic dynamical changes which can be monitored in real time so as to understand the intermediate stages before the system returns to the equilibrium. This is depicted pictorially in Fig. 1 where after excitation by a femtosecond laser pulse (Fig. 1(a)), the system's response R(t) is monitored by an ultrafast probe pulse I(t) through the convolution between the two at each time t (Fig. 1(b)). Material excitations which have life-time broadening larger than the temporal width of the laser pulse can be coherently populated under proper experimental conditions. These coherent states are observed by collecting the time-resolved scattered signal. In Fig. 1(c) we have also shown a situation where due to the ultrahigh peak electric field associated with the

[*]Sunil Kumar
Department of Physics, Indian Institute of Science,
Bangalore 560012, India

[#]Current Address:
Department of Physics, Indian Institute of Technology,
New Delhi 110016, India
Email: kumarsunil@physics.iitd.ac.in

A. K. Sood
Department of Physics, Indian Institute of Science
Bangalore 560012, India

femtosecond laser pulses, one can easily find various possibilities for parametric and nonparametric nonlinear optical processes such as, multi-photon absorption, cascaded two-photon absorption, two-photon excited fluorescence, second or higher harmonic generation, Kerr effect and coherent anti-Stokes scattering.

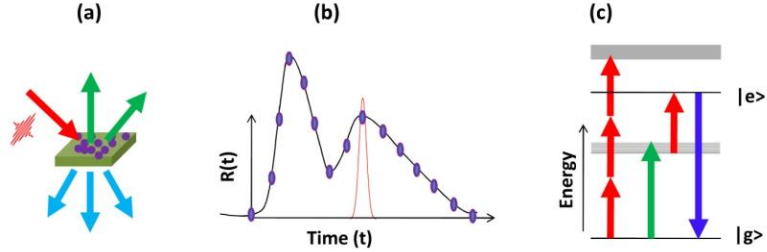

Fig. 1. (a) Photo-excitation of a material system by an ultrashort light pulse. The scattered light can be collected in either time-resolved or spectrally resolved manner using optical schemes which allow detection of smallest possible changes in the system's optical response created by the ultrafast laser pulse and bearing the state character of the dynamical system at a point of observation in the phase space. (b) Following ultrashort photo-excitation, the system evolves through various intermediate dynamical states which can be monitored by using an ultrashort optical probe pulse that measures the convolution between the system's response function and the probe itself. (c) The ultrahigh peak electric fields associated with the ultrashort laser pulses induce various possible nonlinear processes in the system such as three-photon absorption, cascaded two-photon absorption and two-photon excited fluorescence, to name a few which can be easily detected.

Plasmons are coupled modes of electromagnetic field of the light and the collective oscillations of conduction electrons in metals. Also, the coupled electromagnetic fields are confined only within a thin skin layer on the surface of the metal leading to a large enhancement of the local electromagnetic field. Surface plasmons can be localized, for example, in nanoscale particles with large surface to volume ratio, or propagating, such as in interfaces between metals and dielectrics. Propagating surface plasmons which need proper optical arrangements for excitation by external light [1-3], have shown potential as a good candidate for future high-speed opto-electro-plasmonic chip scale integrated devices [4, 5]. Similarly, localized surface plasmons in nanoparticles have led to many interesting effects one of which is surface enhanced Raman scattering [6]. Strikingly, the electromagnetic resonance due to the localized surface plasmons in the metal nanoparticles is easily tunable by shape, size and choice of the surrounding dielectric medium, and occurs in the visible to near infrared region of the electromagnetic spectrum. The plasmons in nanoparticles can be directly excited by incident light and lead to strong light scattering and absorption due to large local optical field enhancements inside and in the neighbourhood of the particles. These appealing characters of metal nanoparticles make them very attractive from the point of view of fundamental studies as well as usability in various technological applications [7-11]. The surface confinement effects in metal nanoparticles cause additional intrinsic dynamical processes which get themselves involved in determining the overall response to an external stimuli such as a femtosecond laser pulse.

A very good understanding of the linear optical properties of plasmonic nanostructures can be achieved simply by applying Maxwell's classical theory of electromagnetism with appropriate boundary conditions and continuity equations at the interfaces between the metal and the surrounding medium and the other intrinsic and extrinsic size effects. The readers can find very good literature available on the synthesis, theoretical and experimental investigations of the linear optical properties of metal nanostructures [10-15]. In this chapter, we focus on the ultrafast response of plasmonic nanostructures, specifically the noble metal nanoparticles. We will discuss the optical nonlinearities and electron dynamics after ultrashort photo-excitation. The emphasis will be on these effects near the plasmonic resonance and away from it. The consequences of reduced dimensions in nanoparticles are not only related to the electronic confinement but also in other attributes such as the confined surface acoustic phonons. These excitations have unique signatures of the thermal and elastic properties of the material and have been studied for quite a long time using Raman scattering, infrared absorption and terahertz time-domain spectroscopy techniques. We will discuss some of these observations as well. Metal nanoparticles with only a few atoms termed as nanoclusters make another interesting area of active research due to very interesting optical phenomena observed in them [16-18]. Such metallic systems of sub-nanometer size lack surface plasmon resonance, the typical feature of bigger nanoparticles in the size range of ~2-100 nm. Instead they show characteristic peaks in the linear absorption spectrum which are signatures of their molecular nature. In this chapter, we have also discussed experimentally observed ultrafast nonlinearities in 15-atom gold clusters where it was seen that the optical limiting performance of the nanoclusters is many folds enhanced when placed in contact with a thin metallic film of indium tin oxide.

    The chapter has been organized as follows. We begin with a brief introduction in Section 2 about the linear optical properties of metal nanoparticles which are essential to understand the ultrafast responses presented in subsequent sections. The absorption and scattering by nanoparticles is discussed in the light of Mie's theory [19] and Gans approximation [20]. We will consider mainly spherical and cylindrical nanoparticles to bring the notions of shape-dependence of the localized surface plasmon resonance. Since in this chapter our focus is on metal nanoparticles, the surface plasmon resonance means the localized surface plasmon resonance. In Section 3, we have discussed the electronic relaxation dynamics in metal nanoparticles following ultrafast optical-excitation at the surface plasmon resonance and away from it. Theoretical models for understanding the underlying relaxation mechanisms have been discussed. We should note that most of the experimental studies on metal nanoparticles have been performed and discussed for inhomogeneous distributions of the particles around a mean value of their size and hence ensemble averaging should be assumed unless mentioned otherwise. Single particle optical spectroscopy measurements are generally more sophisticated in terms of experimental realizations, some of which will be mentioned while discussing the ultrafast time-resolved response in Section 3. In Section 4, we will describe the ultrafast nonlinear optical response of metal nanoparticles. For a flavour of nonlinear optical response from subnanometric and nonplasmonic particles, we will discuss an interesting example of 15-atom gold clusters which show an improved optical limiting performance at wavelengths near the surface plasmon resonance of bigger nanoparticles which otherwise would have shown saturable absorption. Section 5 discusses effects of phonon confinement in nanoparticles measured by the ultrafast optical response from them. Experimental studies reporting the characterization of the confined acoustic phonons in spherical and rod-shaped nanoparticles will be discussed. Finally we will conclude by providing an assessment of the potential opportunities and challenges for metal nanoparticle plasmonics in Section 6.

## 2    Surface plasmons in metal nanoparticles

The optical properties of materials can be completely described by the complex dielectric function ($\varepsilon = \varepsilon_1 + i\varepsilon_2$) or complex index of refraction ($\tilde{n} = \varepsilon^{1/2}$). The dielectric function of metals consists of contributions from the free electrons or the conduction band electrons (intraband transitions) through the Drude term $\varepsilon^D$ and the bound electrons from inner electronic bands (interband transitions such as by d-band electrons in noble metals) through a Lorentzian term $\varepsilon^{ib}$,

$$\varepsilon(\omega) = \varepsilon^D + \varepsilon^{ib} = 1 - \frac{\omega_p^2}{\omega^2 + i\Gamma_0 \omega} + \frac{\tilde{\omega}_p^2}{\omega_0^2 - \omega^2 - i\gamma\omega} \tag{1}$$

Here $\omega_p$ is the lowest cut-off frequency of the collective motion of the free electrons inside metals termed as the bulk plasma frequency, given by $\omega_p = (n_e e^2 / \varepsilon_0 m_e)^{1/2}$ for electronic density $n_e$, charge $e$, mass $m_e$ and vacuum permeability $\varepsilon_0$. The parameter $\Gamma_0$ is the free-electron scattering rate, $\tilde{\omega}_p$ and $\gamma$ are analogous plasma frequency and scattering rate for the bound electrons with interband transition energy at $\hbar\omega_0$. Neglecting the interband contribution and at high frequencies ($\omega$) such that $\Gamma_0/\omega \ll 1$, Eq. (1) provides the more commonly used expression, $\varepsilon(\omega) = 1 - \omega_p^2 / \omega^2$. The plasma frequency ranges between ~1 to 6 eV in various metals below which they do not absorb light and this is the reason why metals are high reflectors in the visible and near infrared range. As the physical dimension of metal structures is reduced to be comparable to the de Broglie wavelength, i.e., of the order of the electron scattering length in metals, confinement effects become dominant leading to significantly different optical properties of metal nanostructure than the bulk metals. For large surface to volume ratios, the plasmons become localized due to the surface effects. The shape, size and composition of the nanoparticles as well as the surrounding medium in which they are embedded dictate the interaction of light with the nanoparticles. Below, we discuss the cases of spherical and cylindrical nanoparticles which are simplest in terms of obtaining analytical solutions for the extinction cross-section.

Gustav Mie [19] for the first time analytically solved the problem of scattering and absorption by a spherical metal nanoparticle. This allows theoretically predict the linear response of a metal sphere to an external electromagnetic field by incorporating the material dielectric properties and solving the Maxwell's equations under appropriate boundary conditions. Consider a spherical particle of radius/volume R/V and metal dielectric function $\varepsilon_m$ that is embedded in a dielectric medium with dielectric function $\varepsilon_d$. For $R$ much smaller than the wavelength of light such that the particle can be assumed to be an ideal dipole in the applied field, Mie's theory predicts the absorption (abs) and scattering (sca) cross-sections [12, 13] to be:

$$\sigma_{abs}(\omega) = \frac{k_d}{\varepsilon_0 \varepsilon_d} \text{Im}(\alpha) \tag{2a}$$

$$\sigma_{sca}(\omega) = \frac{k_d^4}{6\pi(\varepsilon_0 \varepsilon_d)^2} (\alpha)^2 \tag{2b}$$

Here, $k_d$ is the light wave-vector in the dielectric medium and $\alpha$ is the linear polarizability given by the Clasusius-Mosotti relation:

$$\alpha(\omega) = 3V\varepsilon_0\varepsilon_d \frac{\varepsilon_m - \varepsilon_d}{\varepsilon_m + 2\varepsilon_d} \qquad (3)$$

It is clear that surface plasmon resonance in spherical nanoparticles is observed when the condition $\varepsilon_m = -2\varepsilon_d$ is satisfied.

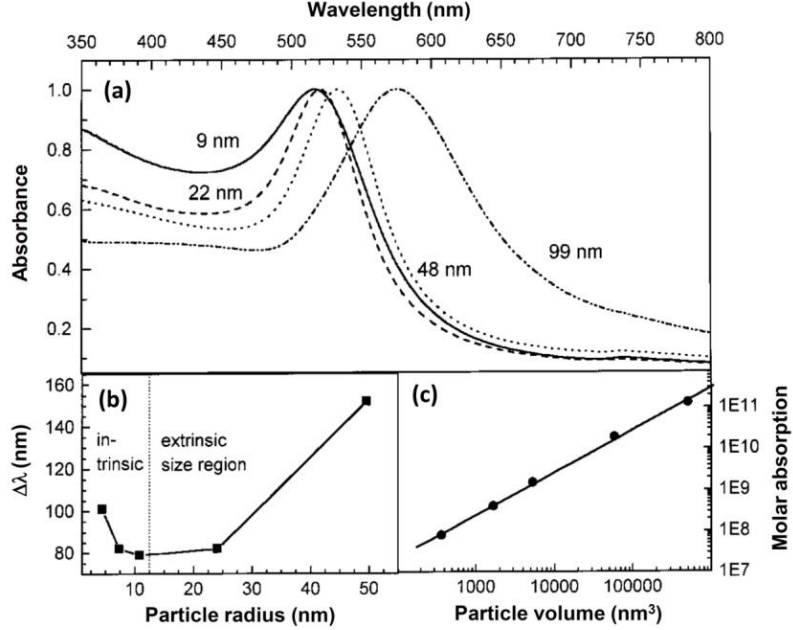

Fig. 2. (a) UV-Visible absorption spectra of colloidal solutions of spherical gold nanoparticles with diameters varying between 9 and 99 nm, (b) plasmon bandwidth as a function of the particle radius, and (c) the extinction coefficient at the respective plasmon absorption maxima plotted against their volume on a log-log scale. The solid line in (c) is a linear fit illustrating the agreement with the Mie theory. Adapted from reference [22] with permission.

In the second example, we consider rod-like nanostructures with dimensions still much smaller than the wavelength of light. For small ellipsoidal particles, scattering problem was solved by Gans in 1912 [20], according to which, the linear polarizability along one of the axes ($k = 1,2,3$) is given by [12, 13, 15]

$$\alpha_k(\omega) = V\varepsilon_0\varepsilon_d \frac{\varepsilon_m - \varepsilon_d}{g_k(\varepsilon_m - \varepsilon_d) + \varepsilon_d} \qquad (4)$$

Here $g_k$ are geometric factors along the three axes of the ellipsoid. For a particular case of spheroidal particles for which the two axes of the ellipsoid are equal and considering $g_1$ along the long axis (a > b = c), the geometrical factors can be expressed as:

$$g_1 = (1/e^2 - 1)\left[(1/2e)\ln\left(\frac{1+e}{1-e}\right) - 1\right]; \quad g_2 = g_3 = (1 - g_1)/2 \qquad (5)$$

where $e = \sqrt{1 - b^2/a^2}$ is the eccentricity of the particle related to the particle aspect ratio $\zeta$ (length/width = $a/b$). Equation (4) in combination with Eq. (2) is used to calculate the scattering and absorption cross-sections for nanorod structures. Clearly there are two resonances in the spectrum for the nanorods, one due to the electron oscillations along the diameter (short axis) called the transverse surface plasmon (TSP) resonance and the other along the length (long axis) of the particles called the longitudinal surface plasmon (LSP) resonance. The LSP resonance is tunable by varying the aspect ratio $\zeta$ but it has significantly small effect on the TSP resonance. Also, the cross-sections are larger for bigger aspect ratios simply because of increase in the volume of the particle. For randomly oriented ensemble of nanorods, the cross-sections are independent of the light polarization and one can use the average quantities, i.e., $\langle\alpha\rangle = (\alpha_1 + \alpha_2 + \alpha_3)/3$ and $\langle\alpha^2\rangle = (\alpha_1^2 + \alpha_2^2 + \alpha_3^2)/3$.

One may take note of few things. Firstly, for small particles (size << wavelength), the scattering is negligible and the experimentally measured extinction cross-section is dominated by the absorption cross-section. Secondly, not only the size of the particles but also the type of the surrounding dielectric medium can be varied for continuously tuning the surface plasmon resonance. We further note that the bulk dielectric functions $\varepsilon_m$ and $\varepsilon_d$ are used to calculate the nanoparticle scattering and absorption cross-sections. In the literature, experimental data by Johnson and Christy [21] on the spectral dielectric function of noble metals are considered as bench marks.

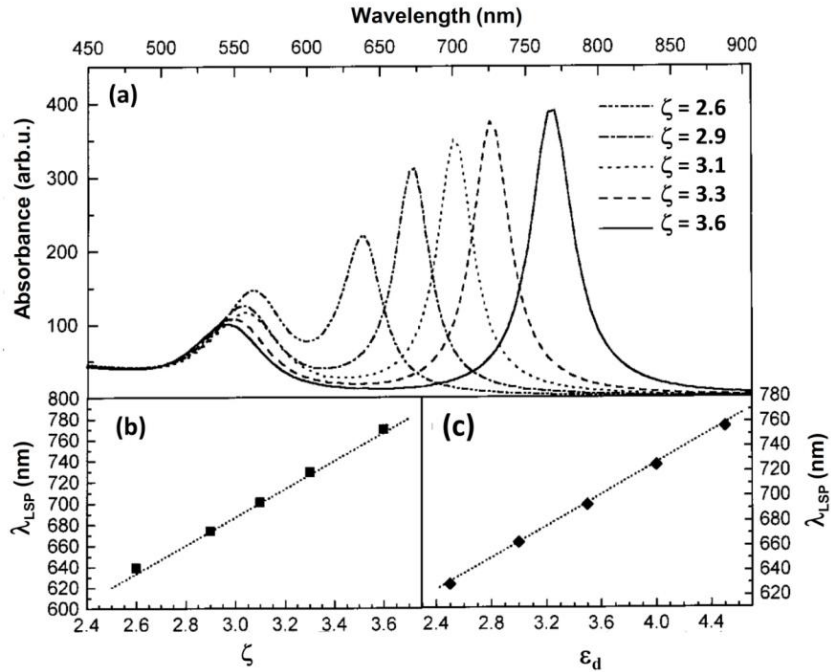

Fig. 3. (a) Tunability of the surface plasmon resonance in gold nanorods simulated as a function of the nanorod aspect ratio $\zeta$. Linear-dependence of the longitudinal surface plasmon resonance wavelength $\lambda_{LSP}$ on,(b) the nanorod aspect ratio for a fixed $\varepsilon_d = 4$, and (c) the dielectric constant $\varepsilon_d$ of the surrounding material for a fixed $\zeta = 3.3$. Adapted from reference [22] with permission.

The tunability of the surface plasmon resonance due to the finite size-effects in spherical gold (Au) nanoparticles and cylindrical gold nanorods [22] is summarized in Figs. 2 and 3, respectively. The red-shift in the surface plasmon absorption maximum as a function of the size of the gold nanospheres is shown in Fig. 2(a) while the dependence of the plasmon bandwidth ($\Delta\lambda$) on the particle radius is shown in Fig. 2(b). Is can be seen that the bandwidth increases with decreasing nanoparticle radius in the intrinsic size region (<10 nm) and also with increasing radius in the extrinsic size region as predicted by the Mie theory. In Fig. 2(c), the extinction coefficients of these gold nanoparticles at their respective plasmon absorption maxima are plotted against their volume on log-log scale where the solid line is a linear fit to the data illustrating an agreement with the Mie theory. Calculated absorption spectra of gold nanorods of different aspect ratio $\zeta$ using the theory developed by Gans are presented in Fig. 3(a). The linear-dependence of the LSP wavelength $\lambda_{LSP}$ on the gold nanorod aspect ratio $\zeta$ for a fixed $\varepsilon_d = 4$ and as a function of the medium dielectric constant $\varepsilon_d$ for fixed $\zeta = 3.3$ can be seen from Figs. 3(b) and 3(c), respectively.

In the Drude model of free electron metals, the only scattering that an electron undergoes is with the other electrons inside the metal. However, for small nanoparticles with large surface to volume ratio, the electron surface scattering and size-dependent radiation damping [23] have to be taken into account. In order to satisfy the energy momentum conservation in the process of a photon being absorbed by a free electron, participation of an auxiliary electron or phonon or a defect is implicit. Therefore, the scattering rate $\Gamma_0$ in Eq. (1) has to be replaced by an effective scattering rate parameter $\Gamma_{eff}$ which is dependent on the incident photon with angular frequency $\omega$, electronic and lattice temperature ($T_e$ and $T_L$, respectively) and contains atleast the additional surface scattering term. Ignoring the scattering from defects, the effective scattering rate [24] can be written as:

$$\Gamma_{eff}(\omega, T_e, T_L) = \Gamma_{e-ph}(\omega, T_e, T_L) + \Gamma_{e-e}(\omega, T_e) + \Gamma_{e-S}(\omega, T_e) \qquad (6)$$

The electron-phonon damping term ($\Gamma_{e-ph}$) is the dominant one which is computed by modelling the electron-phonon interaction *via* deformational potential connecting the bottom of the conduction band to the periodic deformation induced by lattice vibrations. Since phonons are much lower energy excitations than electrons, electron-phonon interaction modifies only the momentum of the excited electrons. The second term in Eq. (6), $\Gamma_{e-e}$ describes the electron-electron scattering which is possible only by exchange of a reciprocal lattice vector *via* Umklapp processes and hence its contribution in the photon absorption by the electron is small.

The last term, $\Gamma_{e-S}$ describes the electron scattering by surface, i.e., optical absorption assisted by electron-surface scattering. It is very significant for confined systems such as nanoparticles of size in the range of ≤30nm, the mean free path of conduction electrons in metals. An empirical relation for the surface scattering term [24] is given as

$$\Gamma_{e-S}(\omega, T_e) = g(\omega, T_e) \upsilon_F/D \qquad (7)$$

where $\upsilon_F$ is the Fermi velocity of electrons and $D = (S_p/\pi)^{1/2}$ is the equivalent diameter of the particle with surface area $S_p$. $g$ is another material parameter which depends on the electron level occupation number defined for quasi-continuous conduction band states of not too small particles.

# 3    Ultrafast optical response of photoexcited metal nanoparticles

The surface confinement at the nanometric scale affects the internal dynamical processes in metal nanoparticles which modify the dielectric function and hence their optical response. As a consequence, after optical excitation by an ultrafast laser pulse, the electron thermalization, energy loss by interactions with mechanical degrees of freedom in the system and energy transfer by heat dissipation to the surrounding; all are modified as compared to those in the corresponding bulk metal. Time-resolved pump-probe spectroscopy offers a mean to investigate the internal dynamical processes by measuring the time evolution of the ultrafast optical response. More specifically, a strong pump pulse creates a dynamical state whose evolution in time is monitored by another weak probe pulse by collecting the scattered signal from the sample as a function of the time-delay between the pump and the probe pulses. As depicted in Fig. 4(a), the routine pump-probe measurements are performed on the ensemble of nanoparticles [22, 25-27], however, experiments on single nanoparticles have also become possible recently [28-32]. Following the ultrafast photo-excitation by a femtosecond laser pulse, the laser deposited energy in the system under investigation relaxes via various possible pathways [24], for example, initial relaxation by electron gas thermalization followed by electron-lattice interaction and finally emission of acoustic vibrations and heat loss to the surroundings, as depicted in Fig. 4(b).

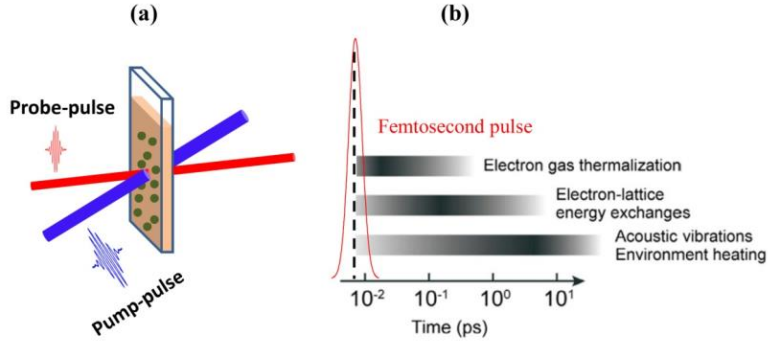

Fig. 4. (a) Pump-probe spectroscopy of an ensemble of nanoparticles arranged for measurement of the transient differential transmission spectra of the probe. (b) Typical time-scales of dynamical processes undergoing in photo-excited metal nanoparticles.

An understanding of the underlying internal system kinetics from the experimentally measured time-resolved optical response can be made based on models which take into account the instantaneous time-dependent modifications of the material dielectric function in the desired spectral range. Detailed quantitative modelling have been performed by Fatti and Vallee [24] which include the impact of the out-of-equilibrium conditions (electron excitation and subsequent lattice heating) on the electron energy distribution, the induced modifications of the metal dielectric function, and their influence on the nanoparticle optical response. A brief overview of the same is presented here as the following.

For nanoparticles in size range of ~2-100 nm we can neglect the scattering losses, i.e., the extinction cross-section to be simply the absorption cross-section $\sigma_{abs}$. Ultrafast optical excitation of a nanoparticle leads to time-dependent changes in $\Delta\sigma_{sbs}$ which is directly related to experimentally measured change in the transmission ($\Delta T$) of the probe at wavelength $\lambda$ by the relation:

$$\frac{\Delta T}{T}(\lambda, t) = -\frac{\Delta\sigma_{abs}(\lambda, t)}{S} \tag{8}$$

$S$ being the surface area of the probe laser spot on the sample. For weak pump-induced changes in the nanoparticles [33], the $\Delta\sigma_{abs}$ variation at the probe wavelength $\lambda$ and pump-probe time-delay $t$, can be expressed as:

$$\Delta\sigma_{abs}(\lambda, t) = \frac{\partial \sigma_{abs}}{\partial \varepsilon_1}(\lambda)\Delta\varepsilon_1(\lambda, t) + \frac{\partial \sigma_{abs}}{\partial \varepsilon_2}(\lambda)\Delta\varepsilon_2(\lambda, t)$$

$$= a_1(\lambda)\Delta\varepsilon_1(\lambda, t) + a_2(\lambda)\Delta\varepsilon_2(\lambda, t) \quad (9)$$

$a_{1,2}$ being the stationary spectral-dependent derivatives and $\Delta\varepsilon_{1,2}$ the dynamical spectral-pump-induced variations of the nanoparticle real and imaginary dielectric functions. Therefore, time-resolved signals result from a combination of the dynamical physical effects, which induce variations $\Delta\varepsilon_{1,2}$ and their optical detection, with a spectral sensitivity given by $a_{1,2}$. For all shapes of metal nanoparticles, the localized surface plasmon leads to a strong enhancement of both the linear absorption spectrum and its derivatives $a_{1,2}$ governing the amplitude of the out-of-equilibrium transient response. For spherical gold and silver (Ag) nanoparticles and cylindrical Au nanorods, the calculated spectra [24] are shown in Fig. 5. The dispersion like $a_1$ profile always crosses the horizontal axis near the surface plasmon resonance. This has consequence in determining the nature or polarity of the transient probe signal from pump-probe spectroscopy near the zero time-delay.

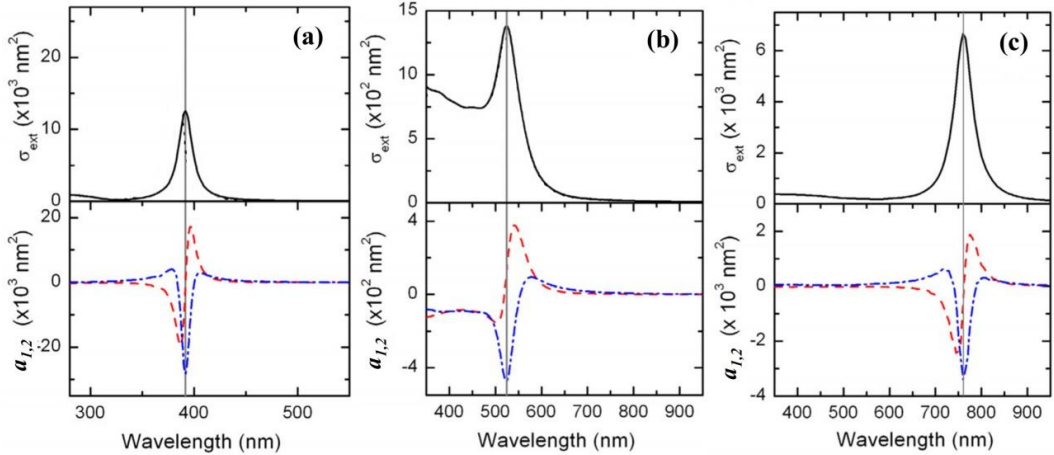

Fig. 5. (Upper panel) Extinction cross-sections, and (lower panel) derivatives $a_1$ (red dashed) and $a_2$ (blue dash-dotted) for a 30 nm diameter silver nanosphere (a), 30 nm gold nanosphere (b) and 43nm × 12nm gold nanorod (c), calculated using Mie theory. The vertical lines indicate the surface plasmon resonance in each case. Adapted from reference [24] with permission.

The time- and energy-dependent distribution function $f(E,t)$ of the electronic states determines the transient variations of the dielectric function $\Delta\varepsilon_{1,2}$ and ultimately the optical response that is measured experimentally. The modification of the electron distribution induces a change in the interband term $\Delta\varepsilon_{1,2}^{ib}$ (dominant on short timescales) and smaller modifications in the Drude term $\Delta\varepsilon_{1,2}^{D}$, and the concomitant heating of the lattice at much longer time-scales. After the pump-pulse excitation of the nanoparticles, the evolution of the electron occupation number, electron-electron scattering and electron energy transfer to the lattice through electron-phonon scattering and heat dissipation to the surrounding medium, all can be computed by solving the Boltzmann equation provided a size-dependent correction to the electron-electron and electron-phonon coupling rates has been taken into account [24]. For pump-photon energy lower than the interband transition energy, electron-hole pairs are generated only in the conduction band. Electrons are thus described by an athermal distribution with a small fraction of electrons having absorbed a photon promoted to a state above the Fermi energy while most of the electrons still occupying initial unperturbed states. Immediately after excitation by a short femtosecond laser pulse, the amplitude of the excitation can be quantified by an elevated electron temperature $T_{exc} = T_0 + \Delta T_{exc}$, $T_0$ being the initial temperature of thermalized electrons before the optical excitation, corresponding to the equilibrium temperature of an electron gas with the same total energy as the excited system. The electron gas thermalizes through internal electron-electron interactions which is described by a screened Coulomb interaction potential containing a sum over all possible two-electron scattering processes satisfying energy and momentum conservation. In low perturbation regime, typical internal thermalization timescales are of the order of ~500 fs. A deformation potential coupling is assumed to compute the electron-phonon coupling matrix element which is then integrated over all available electronic and phonon states satisfying energy and momentum conservation to obtain the contribution from electron-phonon scattering. The time-scale for electron-phonon scattering processes is typically in the picosecond range and depends on the excitation conditions.

Following the internal thermalization by electron-electron scattering, the subsequent dynamics of the electrons and phonons can be described by the two temperature model (TTM) which assumes that both the conduction electrons and the phonons are internally thermalized to have attained a constant coupling term $G$ [34]. Assuming $T_e$ and $T_L$ as the temperatures of the internally thermalized electrons and lattice, and $C_e$ and $C_L$ as their specific heats per unit volume, the system dynamics follows a coupled rate equation given by

$$C_e \frac{dT_e}{dt} = -G(T_e - T_L); \qquad C_L \frac{dT_L}{dt} = G(T_e - T_L) \qquad (10)$$

The final equilibrium temperature $T_{eq}$ of the nanoparticles common to the electron gas and the lattice, generally much smaller than $T_{exc}$ is determined from the analytical solution as $T_{eq} = T_0 + (T_{exc}^2 - T_0^2)/(2C_L/a)$ where $a = 65$ J/m$^3$/K$^2$ for silver and gold [35]. Since the electron-lattice temperature $T_{eq}$ is higher than the initial temperature $T_0$, energy is subsequently transferred to the surrounding medium through the nanoparticle interface. The heat transfer out of the nanoparticles occurs typically with 10 to 500 ps timescales and is limited by the thermal impedance at the interface and by the thermal diffusion within the surrounding medium [36].

First time-resolved experimental results on a single 30 nm silver nanosphere obtained by Muskens et al., [29] using 140 fs pump excitation at 850 nm are reproduced in Fig. 6. The probe pulses were close to the surface plasmon resonance at ~425 nm. The transient differential transmission ΔT/T data for different pump powers as presented in Fig. 6(a) show a fast rise followed by a slower decay corresponding to energy injection into the electron gas by the pump pulse and subsequent electron energy loss by thermalization with the lattice, respectively. Experiments on single nanoparticles allow precise measurement of the $\Delta T_{exc}$ which is not the case for optically excited ensemble of nanoparticles. Dotted, dashed and solid lines in Fig. 6(a) are for 180, 280 and 480 μW pump power, corresponding to $\Delta T_{exc} =$ 190, 275 and 415 K, respectively. Figure 6(b) shows the simulated signals computed using Eq. (9) for which the derivatives $a_1$ and $a_2$ were determined from fitting the measured extinction cross-section by Mie theory (shown in the inset). The ultrafast kinetics is mainly governed by the time-dependent $\Delta\varepsilon_1^{ib}$ while $\Delta\varepsilon_2$ contribution is negligible as the probe-photon energy is far from the interband transition energy in silver.

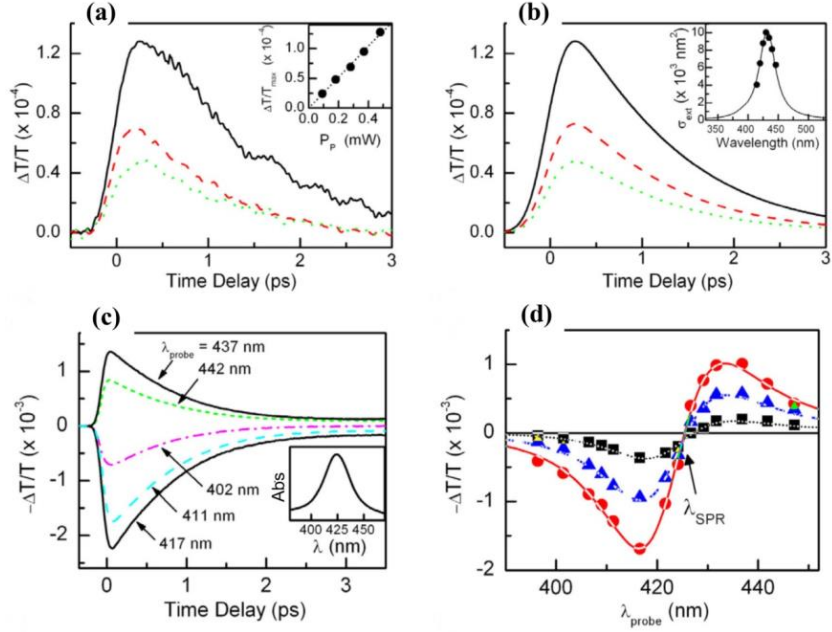

Fig. 6. (a) Transient differential transmission spectra measured for a single 30 nm Ag nanosphere supported on a glass substrate, following excitation by 140 fs pump pulses centred at 850 nm and probe pulses centred at 425 nm near the surface plasmon resonance of the nanoparticle. Dotted, dashed and solid lines correspond to 180, 280 and 480 µW pump power, corresponding to $\Delta T_{exc}$ = 190, 275 and 415 K, respectively. The inset shows the maximum of the $\Delta T/T$ signal as a function of pump power. (b) Simulated signals computed with Eq. (9), $a_1$ and $a_2$ derivatives being determined after fitting the measured extinction cross-section by Mie theory as shown in the inset. (c) Similar experimental results on ensemble of silver nanospheres of average diameter 26 nm dispersed in a glass matrix using 30 fs pump pulses cantered at 850 nm and variable probe pulses between 400 and 450 nm. The inset shows the sample absorption spectrum. (d) Probe-wavelength dependence of the sample transmission change for probe delay of 400 fs (circles), 1 ps (triangles) and 2 ps (squares). Lines are fits assuming pump-induced SPR frequency shift and the broadening. Adapted from reference [24] with permission.

More conventionally, time-resolved experiments have been performed on nanoparticle ensembles, either embedded in a dielectric matrix or dispersed in a liquid. In this case, due to the inhomogeneous distributions of shapes and sizes, and polarization-independent pump absorption, the measurements are less quantitative; however, the intrinsic dynamical characteristics of the nanosystems are captured. Neglecting the absorption by the surrounding matrix, for an ensemble of nanoparticles within an effective optical length $l_{op}$, the differential transmission signal of the probe can be written as:

$$\frac{\Delta T}{T}(\lambda, t) \sim -\Delta\alpha(\lambda, t).l_{op} \sim -n_{np}\Delta\bar{\sigma}_{abs}(\lambda, t).l_{op} \qquad (11)$$

where $\Delta\bar{\sigma}_{abs}$ is the mean value of the absorption cross-section of the ensemble with the volume density $n_{np}$ of the nanoparticles. Results from one of the earliest investigations on silver nanospheres [37] are presented in Fig. 6(c) where 30 fs pump pulses at 850 nm and probe pulses varying between 400 to 450 nm were used to study silver nanospheres of mean diameter 26 nm dispersed in a glass matrix. The dispersion-like dependence on the probe-wavelength around the surface plasmon resonance at 425 nm is shown in Fig. 6(d) for probe delay of 400 fs (circles), 1 ps (triangles) and 2 ps (squares). Clearly, the nature of the transient response of the nanoparticles near zero probe-delay depends on the relative position of the probe wavelength with respect to the surface plasmon resonance as a consequence of the pump-induced transient frequency-shift and -broadening of the SPR. Following the pump-excitation, the $\Delta\varepsilon_1^{ib}$ increases very fast leading to a red-shift of the SPR and then decays to its initial value within a few ps by electron-lattice coupling. The SPR broadening at ultrafast time-scales arises due to change in $\Delta\varepsilon_2^{ib}$ modifying the state occupation number during and immediately after the excitation, and at slower time-scales due to changes in the strength of the scattering processes.

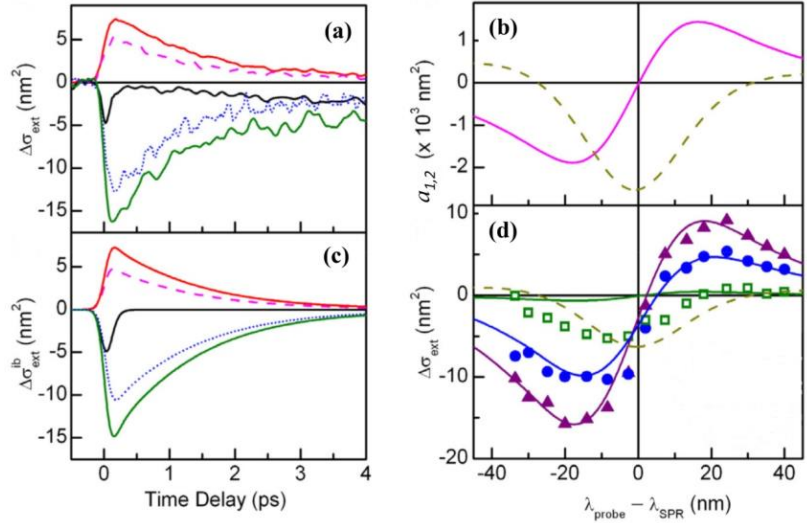

Fig. 7. (a) Ultrafast extinction or absorption cross-section changes $\Delta\sigma_{ext}$ measured for a single 43nm ×12nm Au nanorod following pump-excitation at 400 nm and probed using pulses with varying wavelengths around nanorod LSP resonance ($\lambda_{LSP}$ = 810 nm): from top to bottom, $\lambda_{probe}-\lambda_{LSP}$ = 30, 40, 0, –30 and –20 nm. (b) Computed interband contributions (Eq. (9)). (c) $a_1$ and $a_2$ coefficients (solid and dashed lines, respectively) computed from the linear extinction cross-section. Vertical line corresponds to $\lambda_{LSP}$. (d) Measured transient extinction spectra for probe-delay of 0 fs (circles), 200 fs (triangles) and 4 ps (squares), and the corresponding lines are computed from interband contributions. Dashed line represents Drude contribution for the longest delay. Adapted from reference [24] with permission.

For Au nanospheres, the SPR lies in the range of 500 to 550 nm depending on the size and the dielectric environment which overlaps with the onset of the interband transition wavelength of ~650 nm. Selective investigation of the SPR dynamics requires spectral separation of the SPR and the interband transitions as satisfied for Ag nanospheres and LSP resonance of Au nanorods. Hence the nature of the ultrafast response probed around the SPR of the Ag nanospheres and the LSP resonance of Au nanorods is expected to be similar. Indeed this is the case as seen from Fig. 7 where experimental results from [31] have been presented for a single Au nanorod excited by femtosecond pump pulses at 850 nm and probe pulses with varying wavelength $\lambda_{probe}$ on either side of the LSP resonance at 810 nm far from the interband transition wavelength of gold. The transient response can be interpreted well on the whole timescale as the combination of the dynamical response of the bulk metal amplified by plasmonic effects as shown by simulations of the $\Delta\sigma_{ext}^{ib}$ or the $\Delta\sigma_{abs}^{ib}$ dynamics [24], i.e., variations of $\Delta\sigma_{ext}$ induced by the interband term $\Delta\varepsilon_{1,2}^{ib}$ which reproduce the experiments, both for signal amplitude and time dependence on the first picoseconds after excitation very well. Also transient signals for short delays reflect the $a_1$ shape profile computed from the experimental linear $\sigma_{ext}$ (Figs. 7(c) and 7(d)). This is because $\Delta\varepsilon_1^{ib}$ is approximately undispersed far from $\lambda = \lambda_{ib}$ and $\Delta\varepsilon_2^{ib}$ is small. For $\lambda \approx \lambda_{LSP}$ (black line in Figs. 7(a) and 7(b)), $a_1$ vanishes and $\Delta\sigma_{ext}$ reflects $\Delta\varepsilon_2^{ib}$ dynamics. Conversely, for longer delays ($t \approx 4$ ps), $\Delta\sigma_{ext}$ is not correctly reproduced any more by the $\Delta\varepsilon_{1,2}^{ib}$ terms alone, and inclusion of the $\Delta\varepsilon_2^D$ variation due to lattice heating becomes necessary (experiments and model prediction are represented by squares and dashed line in Fig. 7(d)).

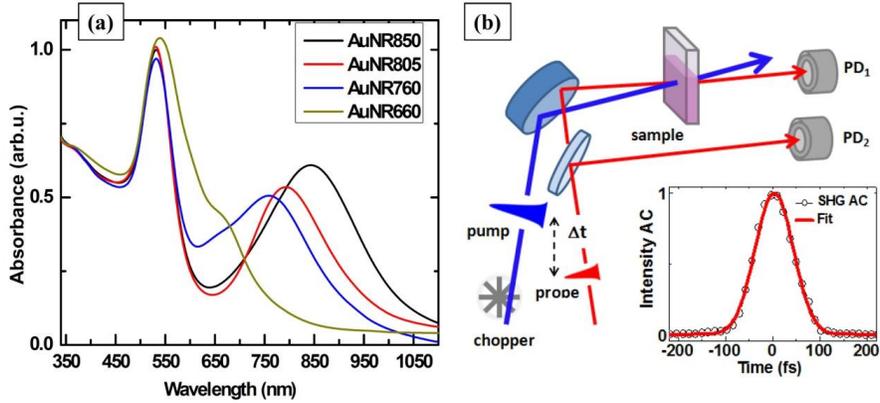

Fig. 8. (a) Linear optical absorption spectra of colloidal suspensions of the gold nanorods in water shown for four representative samples named according to the longitudinal surface plasmon wavelength. (b) Typical experimental arrangement for time-resolved pump-probe spectroscopy. The inset in (b) shows the second harmonic generation intensity autocorrelation (SHG AC) trace of the femtosecond laser pulses centred at 790 nm.

## 3.1 *Tuning between ultrafast PB and PA in gold nanorods by selective probing near LSP resonance*

As discussed above, the nature or polarity of the initial transient response immediately after photo-excitation of metal nanoparticles depends on the probe wavelength relative to the surface plasmon resonance and the condition that the later is spectrally separated from the interband transition wavelength. We present below our experimental observations on gold nanorods where the pump excitation at either wavelength (energy) 790 nm (1.57 eV) or its second harmonic at 395 nm (3.15 eV) creates athermal population of electrons whose relaxation is monitored by probe pulses of fixed wavelength 790 nm. To prove the point, one can either vary the probe wavelength continuously across the LSP resonance of the nanorods or equivalently, use different gold nanorod samples for a fixed probe wavelength. The later has been followed here. In these experiments, transient differential transmission spectra $\Delta T(t)/T$ have been measured where positive change in the probe transmission at the zero probe-delay ($t = 0$) is due to photo-bleaching (PB) of the LSP band and negative change in the probe transmission is due to photo-induced absorption (PA) or excited state absorption induced by the pump of either same wavelength 790 nm (degenerate pump-probe configuration) or different wavelength 395 nm (non-degenerate pump-probe configuration) for both of which the interband transition wavelength of gold ($\lambda_{ib}$ ~ 650 nm) is far off.

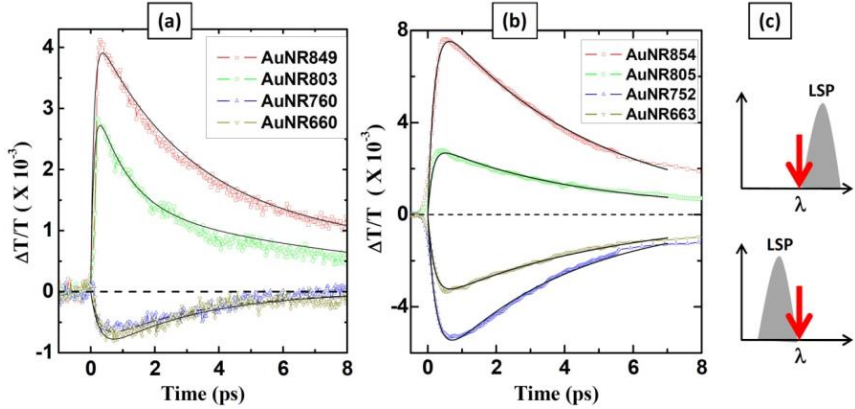

Fig. 9. Transient differential transmission data for gold nanorods taken using (a) degenerate pump-probe spectroscopy at 790 nm, and (b) non-degenerate pump-probe spectroscopy with 395 nm pump and 790 nm probe pulses. Adapted from references [38, 39] with permission. (c) A depiction to emphasize that the ultrafast response near zero probe-delay is sensitive only to the relative position of the probe wavelength (marked by red arrow) with respect to the LSP wavelength.

The linear absorption spectra of four representative gold nanorod colloidal suspensions in water [38, 39], are presented in Fig. 8(a). The gold nanorods have aspect ratio $\zeta$ (length/diameter) ranging from 3 to 5 such that the LSP resonance wavelength (energy) varies from 660 nm to 860 nm ($E_{LSP}$ = 1.88 eV to 1.42 eV). According to the LSP wavelength, the samples are named as AuNR660 to AuNR860 and the same nomenclature will be used in the discussion here. The typical experimental setup used for either degenerated or non-degenerated pump-probe experiments is shown in Fig. 8(b). The differential transmission data ($\Delta T/T$) from four nanorod samples using degenerate pump-probe spectroscopy at 790 nm taken at pump-fluence of ~130$\mu$J/cm$^2$ and probe-fluence of ~8$\mu$J/cm$^2$ are shown in Fig. 9(a). On similar nanorods, results from non-degenerate pump-probe spectroscopy at 395 nm pump and 790 nm probe taken at pump-fluence of ~600$\mu$J/cm$^2$ and probe-fluence of ~1.3$\mu$J/cm$^2$ on are presented in Fig. 9(b). Clearly, PB is observed for samples with $\lambda_{LSP} > \lambda_{probe}$ and PA for samples with $\lambda_{LSP} < \lambda_{probe}$. A depiction in Fig. 9(c) emphasizes the observation that irrespective of the pump-wavelength and pump-flux, the response of the gold nanorods, either the PB (upper diagram) or the PA (lower diagram) is sensitive only to the probe-wavelength relative to the LSP wavelength. The solid curves in Figs. 9(a) and 9(b) are results from numerical simulations assuming a pump-induced time-dependent excess athermal electron density $n(t) = (1 - e^{-t/\tau_R})n_1 e^{-t/\tau}$, $\tau_R$ and $\tau$ being the initial rise time-constant due to thermalization and fast relaxation time-constant, respectively. Essentially, $n(t)$ modifies the instantaneous plasma frequency $\omega_p\{n(t)\}$ and thereby the dielectric constant of the particle, resulting into the observed time-dependence of the differential transmission signal at the probe wavelength for the sample optical path length of $l_{op}$ given by [39],

$$\Delta T(t)/T = -\Delta\alpha\{n(t)\}.l_{op} \qquad (12)$$

## 3.2 *Light controlled reversible switching between ultrafast PB and PA in gold nanorods*

Interestingly, reversible switching between ultrafast PB and PA effects in gold nanorods can also be achieved by using the probe photon-flux as the control parameter [39]. This occurs only for those gold nanorods which have the surface plasmon wavelength above the probe wavelength and the pump wavelength such that it excites the interband transitions. Experimental results obtained on gold nanorods using nondegenerate pump-probe spectroscopy at 395 nm pump-wavelength and 790 nm probe-wavelength, are presented in Fig. 10. For a fixed pump-fluence of ~580 µJ/cm$^2$, the results for AuNR854 are shown in Fig. 10(a) where it can be seen that the amount of the PB signal at low probe-fluences continuously decreases with the increasing probe-fluence and changes to PA at higher probe-fluences. The threshold value of the probe-fluence for which switching from PB to PA occurs is ~80µJ/cm$^2$, independent of the sample and the value of the pump-fluence as can be noted from the results summarized in Figs. 10(b) and 10(c). For the reverse, i.e., decreasing the probe-fluence from high to low, the signal continuously changes from PA to PB without any measurable hysteresis loss. This unique observation is independent of the pump-fluence and occurs even at high pump-fluences but much below the nanorod melting threshold. On the other hand, for the AuNR663 and AuNR752 samples for which the LSP wavelength is below the probe wavelength, we always observe PA as seen in Fig. 10(b) irrespective of the pump or the probe-fluences.

In Fig. 10(c) we have plotted the maximum of the ΔT/T signals near zero-delay as a function of the probe-fluence for the AuNR854 sample. Each curve is for a particular value of the pump-fluence as mentioned. It can be noted that all the curves pass through a common point on the horizontal axis (probe-fluence) irrespective of the value of the pump-fluence and the change in polarity from PB to PA occurs at the probe-fluence of ~80µJ/cm$^2$. If laser heating induced melting of the gold nanorods has to be invoked then it is unusual to have all the curves pass through a single point. Rather the pump and the probe-fluences should have acted in a constructive way to heat up the system additively (function of the total laser-fluence incident on the sample at time delay t = 0). This is, of course, not the case here.

Furthermore, the linear-dependence of the ΔT(t=0)/T on the pump-fluence for a fixed probe-fluence as seen in Fig. 10(d) strengthens the point further that the reversible switching between PB and PA as a function of the probe-fluence is intrinsic to the nanorods and is not due to nanorod shape change due to melting. Our results clearly show that the pump-fluence dependent results at any given fixed value of the probe-fluence are usual, i.e., the magnitude of the transient absorption (either PB at low probe-fluences or PA at high probe-fluences) as shown in Fig. 10 (d), increases linearly with the pump-fluence without change in the polarity.

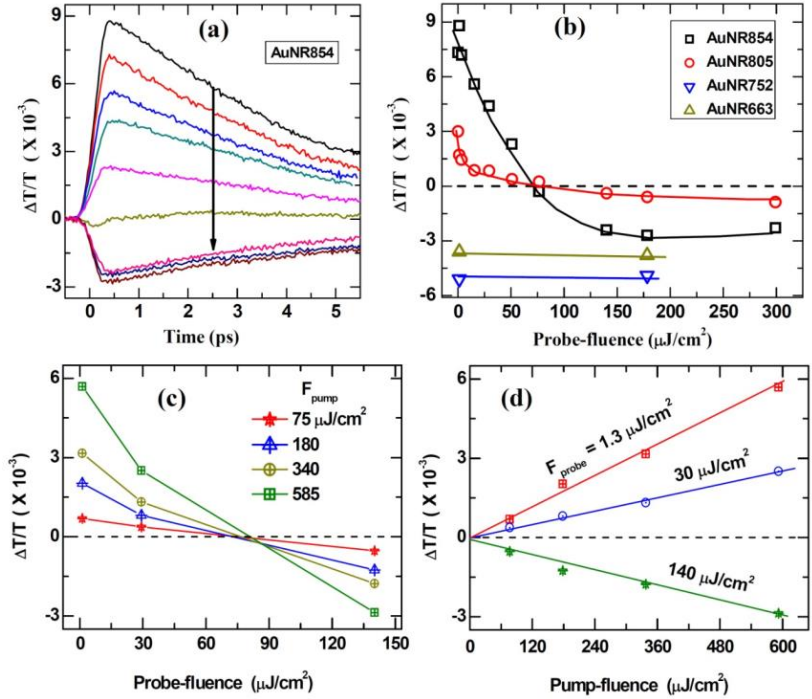

Fig. 10. (a) Transient differential transmission spectra for AuNR854 taken at a fixed pump-fluence of ~580 μJ/cm$^2$ and increasing the probe-fluence in the direction of the arrow. (b) Maximum amplitude of the signals near zero time-delay ($t = 0$) for all the four nanorod samples as a function of the probe-fluence. (c) Probe-fluence dependence of the ΔT(t=0)/T measured at various fixed values of the pump-fluence as mentioned, and (d) probe-fluence dependence of the ΔT(t=0)/T measured at various values of the probe-fluence as mentioned, for the AuNR854 sample. Solid lines in (b), (c) and (d) are guide to the eyes. Adapted from reference [39] with permission.

To explain the observed effect due to the probe-fluence in our experiments, we have to invoke a cascaded two-photon absorption process. The pump at 315 nm (photon energy larger than the TSP and LSP energies) excites electrons from the d-band to tail states of the continuum band above the interband transition energy in gold (~2.4 eV) or the TSP band (~2.7 eV) in the nanorods. The electrons from these excited states quickly thermalize and populate the LSP band within a time $\tau_R$ ~ 300 fs, as reflected in the rise-time of our differential transmission signals at probe wavelength 790 nm, irrespective of the gold nanorod aspect ratio, and the pump and probe fluences. At low values of the probe fluence, the usual photo-bleaching is observed when the pump excited carriers relax to the LSP band and block the normal probe absorption to that band. However, at high probe-fluences, beyond the threshold value of ~80μJ/cm$^2$, the excited state absorption from the LSP band to higher energy states dominates the signal leading to the observed photoinduced absorption.

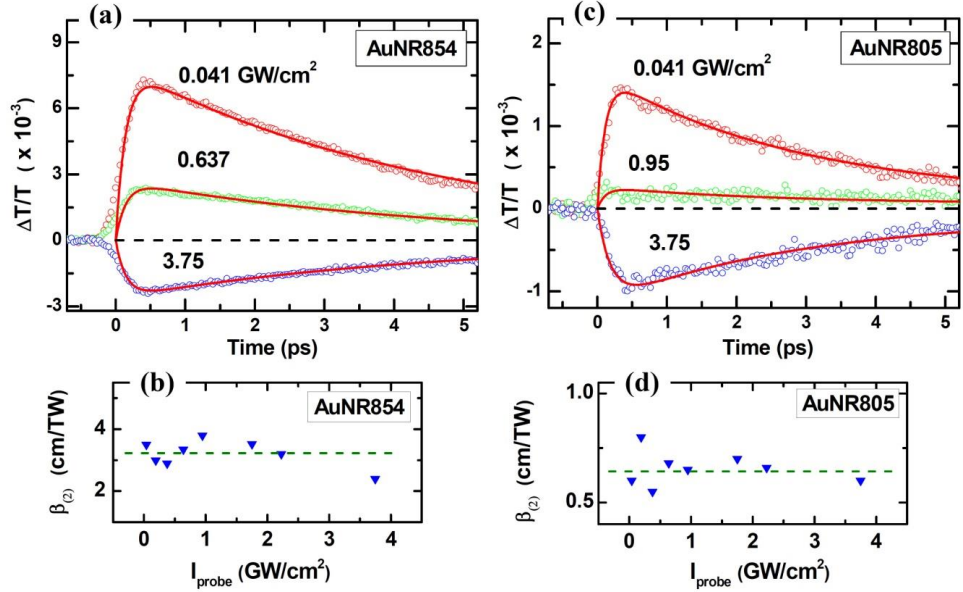

Fig. 11. Transient differential transmission spectra along with the results from simulations (dark curves) shown for (a) AuNR854, and (b) AuNR805 nanorods at a few probe-intensities (GW/cm$^2$) as mentioned after incorporating the cascaded two-photon absorption coefficient $\beta_{(2)}$ at the corresponding probe-intensity from (b) and (d), respectively. Dashed lines in (b) and (d) are guide to the eyes. Adapted from reference [39] with permission.

We consider a special two-photon absorption coefficient $\beta_{(2)}$ which represents the cascaded two-photon absorption, one photon from the pump and one from the probe pushing the electrons from the d-band to the excited states. The action of the pump-fluence is to create an initial excess density $n_1$ of photoexcited carriers which thermalize by electron-electron scattering within time $\tau_R$ and subsequently relax with a time constant $\tau$ by electron-phonon scattering. Assuming that the time-dependence of the contribution arising from the $\beta_{(2)}$ process is the same as before in Eq. (12), i.e., $(1-e^{-t/\tau_R})n_1 e^{-t/\tau}$, the total transmission change is suggested to be of the form:

$$\left(\frac{\Delta T(t)}{T}\right)_{total} = \frac{\Delta T(t)}{T} - \beta_{(2)}(I_{pump} \times I_{probe})^{1/2}(1-e^{-t/\tau_R})n_1 e^{-t/\tau} \qquad (13)$$

where the first term on the right side is same as discussed before in Eq. (12) for the usual case. Using Eq. 13, simulations were carried out to fit the experimental data. We start with a negligibly small $\beta_{(2)}$ at the lowest probe-fluence of 1.3 μJ/cm$^2$ (intensity ~ 16 MW/cm$^2$). Then by adjusting only the value of $\beta_{(2)}$, we simulate the results at various probe-fluences to obtain fits as shown by dark continuous lines in Figs. 11(a) and 11(b). The resultant $\beta_{(2)}$ values as a function of the probe-intensity are shown in Figs. 11(c) and 11(d) for AuNR854 and AuNR805 samples, respectively. We can clearly see from Figs. 11(c) and 11(d) that the resultant $\beta_{(2)}$ is largely independent of the probe-intensity and a simple cascaded two-photon absorption process is critical to explain the observed reversible transition between PB and PA for the AuNR854 and AuNR805 nanorods.

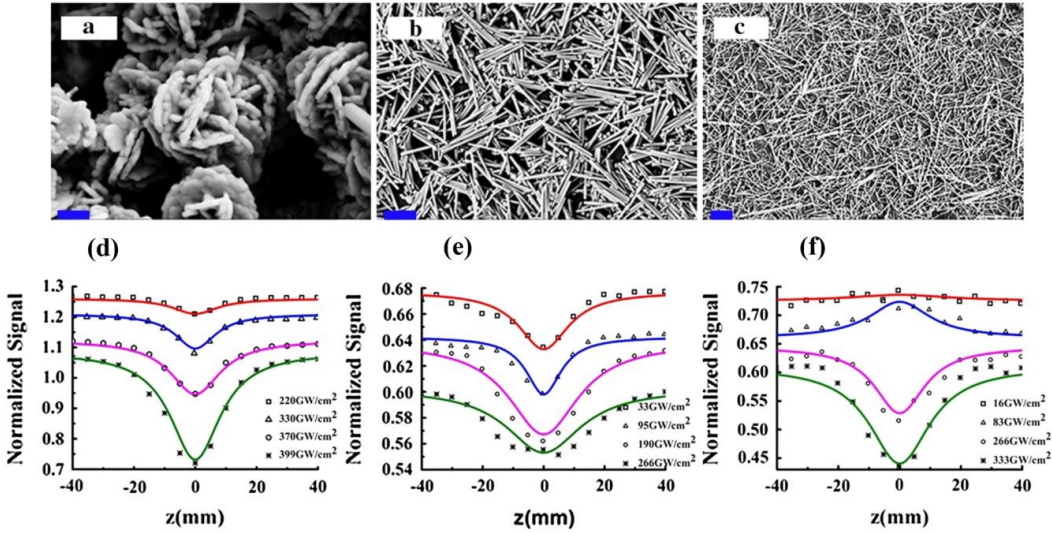

Fig. 12. Scanning electron microscopy images of silver nanoparticles with different shapes, (a) nanoflowers, (b) microrods, and (c) nanowires. The scale bars are 500 nm, 1 micron and 2 microns, respectively. The corresponding nonlinear optical transmission data obtained using open aperture z-scan experiments with 87 fs laser pulses centred at 800 nm, are presented in (c), (d) and (e), respectively. Adapted from reference [46] with permission.

## 4 Ultrafast optical nonlinearities of metal nanoparticles

### *4.1 Surface plasmon resonance tuned optical nonlinearities*

We have already seen that the linear response, i.e., one-photon absorption and scattering cross-sections of metal nanoparticles with sizes typically in the range of ~2-100 nm, are enhanced near the surface plasmon resonance. It is obvious that the surface plasmon resonance in metal nanoparticles would also play a vital role in determining their nonlinear optical properties. It has been seen in many studies that near the surface plasmon resonance, metal nanoparticles exhibit immensely enhanced optical nonlinearities as compared to their bulk counterparts [40-43]. Generally, the optical nonlinearities can be due to electronic transitions or thermal effects. Electronic nonlinearities have ultrafast response, i.e., in the femtosecond to picosecond range while the thermal processes induced nonlinearities are slow, i.e., in the nanosecond or slower time scales. In either case, the nonlinear optical properties of materials can be measured using a single laser beam technique popularly known as the z-scan technique introduced by Sheikh Bahai et al., in 1990 [44]. In the open aperture (OA) configuration, all the transmitted light from a sample is collected onto a photodiode which is a function of the complex optical susceptibility of the material. On the other hand, using close aperture (CA) configuration where only part of the laser beam transmitted through the sample is collected using an aperture before the photodiode, the nature and magnitude of the detected signal depends on the type and magnitude of the real part of the complex optical susceptibility of the sample. Therefore, z-scan technique is useful for measuring the effective cross-sections of various intensity-dependent processes such as second harmonic generation, third harmonic generation, two-photon and multi-photon absorption, and Kerr-effect, to name a few.

The optical nonlinearities in metal nanoparticles are ultrafast in nature, occurring in the range of femtosecond to picoseconds [45]. Very often, the two-photon absorption coefficient β and an intensity-dependent coefficient of refraction γ arising from third-order optical susceptibility $\chi^{(3)}$ are reported in the literature. Nonlinear optical properties of metal nanostructures are closely related to the surface plasmon resonance, i.e., the size and shape of the nanostructures. For silver nanostructures of various shapes, Luo et al., [46] observed that with the increasing incident intensity of 87 fs pulses centred at 800 nm, the reverse saturable absorption or optical limiting behaviour related to $\chi^{(3)}$ increases for nano-size flower-shaped structures but decreases for microrods. However, for silver nanowires, saturable absorption response is observed initially at low incident intensities which changes to reverse saturable absorption with higher nonlinear absorption coefficient at high incident intensities. These results have been reproduced in Fig. 12 where the scanning electron microscopy (SEM) images and the corresponding nonlinear transmission properties of the three samples have been presented.

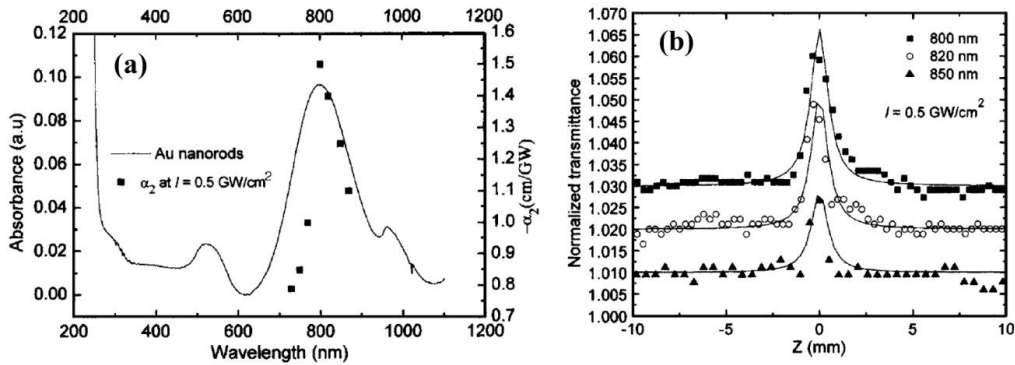

Fig. 13. (a) Spectral-dependence of the linear and nonlinear absorption coefficients of the gold nanorods, (b) open aperture z-scan results on the gold nanorods measured at different excitation wavelengths using 220 fs laser pulses for a fixed incident irradiance of 0.5 GW/cm$^2$. Adapted from reference [47] with permission.

The tunability of the SPR in metal nanostructures makes them unique for various potential applications in biology and medicine. It has been shown in many studies that the optical nonlinearities or the coefficients β and γ measured at the surface plasmon resonance are much enhanced, sometimes by many folds of magnitude as compared to those measured at off-resonance. For example, in gold nanorods, femtosecond z-scan experiments revealed that the saturable absorption behaviour can be continuously changed to optical limiting by increasing the amount of the incident irradiance [47]. These results have been reproduced in Fig. 13. In this study gold nanorods with their TSP resonance at ~520 nm and LSP resonance at ~800 nm were used. The laser excitation wavelength was in resonance with the LSP wavelength. It can be seen from Fig. 13 that the two-photon absorption coefficient ($-\alpha_2$ cm/GW) related to saturable absorption for smaller values of the incident intensity, is maximum for excitation wavelength in resonance with the LSP wavelength.

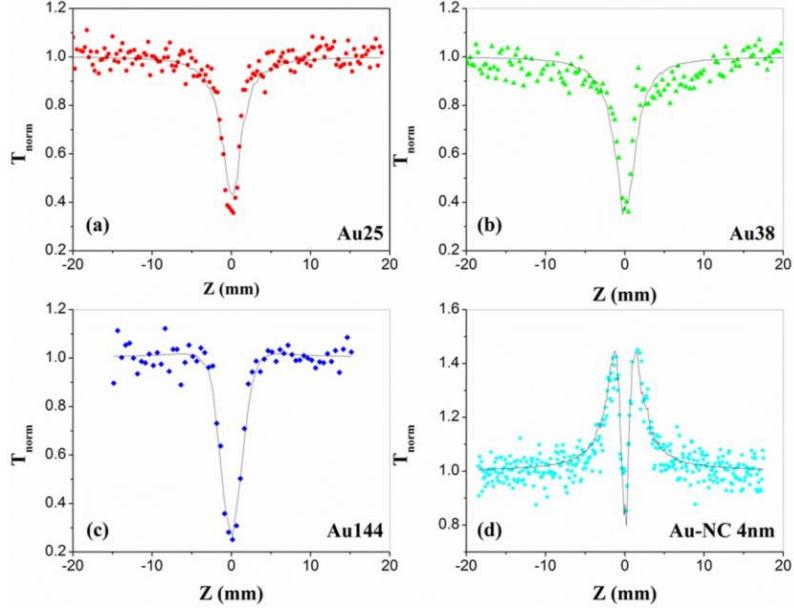

Fig. 14. Open-aperture z-scans measured for gold clusters, (a) Au25, (b) Au38, (c) Au144, and (d) gold nanocrystals of size ~4nm, Au-NC 4nm. Samples are excited using 5 ns laser pulses at 532 nm. $T_{norm}$ is the measured transmission normalized by the linear transmission of the samples. Solid curves correspond to numerical fits to the data. Taken from reference [54] with permission.

## *4.2 Metal nanoclusters with improved ultrafast nonlinearity*

Smaller metal nanoparticles, namely, the nanoclusters containing a few atoms are a new class of materials with characteristics in between those of atoms and nanoparticles [16-18, 48-50]. The nanoclusters with size of the order of the de Broglie wavelength of conduction electrons (∼0.5 nm) exhibit discrete energy levels while larger nanoparticles or nanocrystals (>2 nm) exhibit quasi-continuous electronic bands with the additional surface plasmon resonance as discussed before. The optical absorption spectra of such subnanometric clusters lack the surface plasmon resonance but rather show a distinct absorption onset at the electronic gap between the highest occupied molecular orbital (HOMO) and the lowest unoccupied molecular orbital (LUMO). Noble metal nanoclusters have been shown to exhibit several novel properties [18, 50-52]. Ultrafast photoexcited electron relaxation dynamics in 28-atom gold ($Au_{28}$) nanoclusters were found to show bi-exponential decay with a fast sub-picosecond and another slow nanosecond time-constant, independent of both the laser photon-energy and the pump-fluence [53]. Gold nanoclusters do not show saturable absorption at the surface plasmon wavelength of larger gold nanocrystals [54]. On the other hand, these ultra-small particles exhibit very good optical power limiting performance with significantly reduced threshold for the optical limiting and higher nonlinear optical absorption coefficient.

Open-aperture z-scan experiments by Philip et al., [54] as reproduced in Fig. 14 showed a transition from pure optical limiting behaviour for gold nanoclusters with 25, 38, and 144 atoms to saturable absorption behaviour for nanocrystals of size ~4 nm, all excited with same 5 ns laser pulses centred at 532 nm near the surface plasmon resonance of the nanocrystals. The valley-shaped z-scan curves for the 25- and 38-atom clusters indicate pure optical limiting behaviour due to the nonlinear absorption throughout the z-scan range whereas that of the nanocrystals show a central valley flanked by two symmetric peaks on either side. The onset of the side peaks signifying the onset of saturable absorption is visible in the form of two humps flanking the valley for the 144-atom clusters.

In the following we discuss a case study of 15-atom gold clusters ($Au_{15}$) whose ultrafast nonlinear optical response was found to be immensely enhanced for the clusters deposited on indium-tin-oxide (ITO) metal film as compared with that on $SiO_2$ glass plate [55]. Systematic experiments were carried out using femtosecond z-scan at 395 nm and time-resolved non-degenerate pump-probe spectroscopy (395 nm pump and 790 nm probe) on the gold-clusters deposited on the glass plate ($Au_{15}$-glass) and on the ITO ($Au_{15}$-ITO) as well as separately on bare ITO and glass plates. The pulse-width of the laser pulses was ~80 fs. The optical limiting performance of the $Au_{15}$-ITO system was found to be higher by almost an order than that of the $Au_{15}$-glass system. Concurrently, excited state absorption from transient transmission measurements was also observed to be enhanced by approximately the same order.

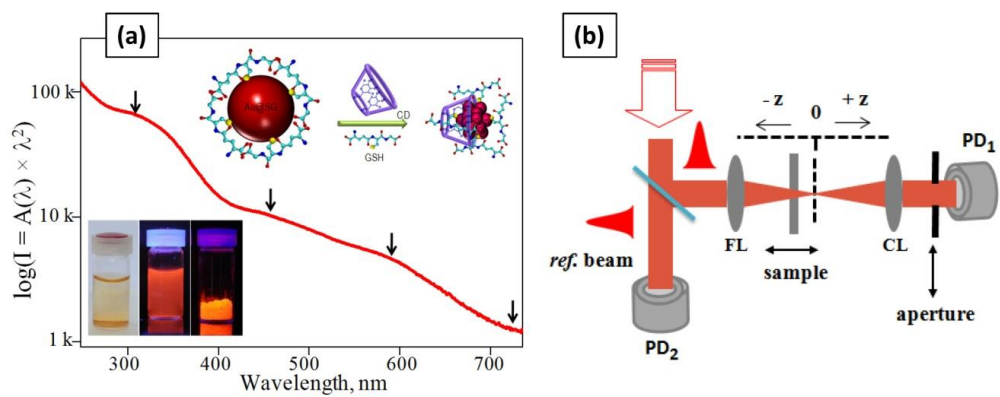

Fig. 15. (a) UV-Vis absorption spectrum of $Au_{15}$ clusters inside cyclodextrin (CD) cavities plotted as the natural logarithm of the Jacobian factor. Well defined absorption features related to molecular character of the clusters are marked with arrows. Lower inset shows fluorescence from the sample in suspension and solid forms. Upper inset shows the schematic illustration of CD-assisted one-pot synthesis of the $Au_{15}$ clusters via the core etching reaction [18]. GSH indicates glutathione used as the core-etching agent and as the ligand to protect the $Au_{15}$ core. Adapted from reference [55]. (b) Experimental setup for open and close aperture z-scan measurements. Convex lenses for focusing the light onto the sample (FL) and collecting the transmitted light from the sample (CL) are used while the sample mounted on a linear stage is moved along the z-axis on both the sides of the focal point z = 0. The transmitted light, either all of it or part of it allowed by an aperture, is detected at a photodiode $PD_1$. Another photodiode $PD_2$ is used as a reference to nullify the noises in the detected signal which are present in the laser beam.

Stable $Au_{15}$ nanoclusters were synthesized involving processes like core-etching of larger clusters and simultaneous trapping of the clusters inside cyclodextrin (CD) cavities [18] shown schematically in the inset of Fig. 15(a). The decorating glutathione (GSH) and CD molecules are optically transparent in the 200-1000 nm window. The absorption spectrum of the gold clusters is shown in Fig. 15(a) where the experimentally obtained wavelength-dependent intensity I($\lambda$) has been converted into energy-dependent intensity I(E) by dividing by the popular Jacobian factor $\partial E/\partial \lambda$. We note that the absorption profile of the clusters does not show surface plasmon resonance (SPR) of bigger nanoparticles. Instead, molecule-like characteristic features marked by arrows in Fig. 15(a) at about 318 nm, 458 nm and 580 nm are clearly visible. The distinct absorption onset occurs near 710 nm which is indicative of an energy gap of 1.75 eV between the highest occupied molecular orbital (HOMO) and the lowest unoccupied molecular orbital (LUMO) gap.

A layer of the $Au_{15}$ nanoclusters embedded in dense hydrogel matrix of thickness ~100 micron was sandwiched between a glass plate (~500 micron thick) and a coverslip (~100 micron thick) making the $Au_{15}$-$SiO_2$ sample and another between an ITO thin film-coated glass plate and a cover slip to make the $Au_{15}$-ITO sample. The commercial ITO plates having film thickness of ~100 nm and a sheet resistance of ~10 ohm per square area were used. When excited with 395 nm laser pulses, a yellowish glow is seen at the back of the samples, which was rejected by using a colour glass filter from reaching the photo-detector in all the experiments described here.

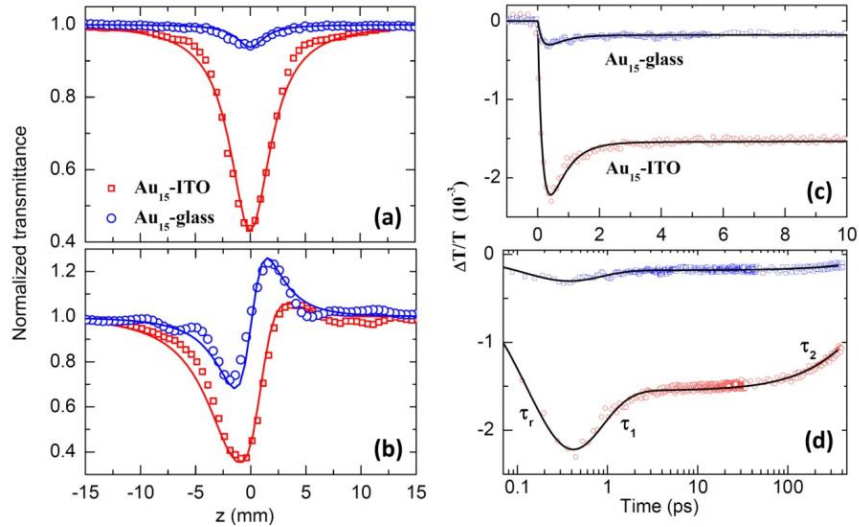

Fig. 16. Normalized transmittance of the $Au_{15}$ clusters in (a) OA z-scan and (b) CA z-scan, shown for clusters deposited on ITO plate (sample $Au_{15}$-ITO) and on glass plate (sample $Au_{15}$-glass) having same molar concentration and layer thickness taken at excitation wavelength of 395 nm and maximum incident power of ~70 GW/cm$^2$ (at the focus). The continuous lines are numerical fits to determine nonlinear absorption and refraction coefficients. Time-resolved differential transmission data for the two samples shown on linear-linear scale (c) and log-linear scale (d), taken using 395 nm pump and 790 nm probe pulses to show the difference in the signal magnitude for the two samples and the related carrier relaxation dynamics involving initial thermalization time-constant $\tau_r$, fast and slow decay time-constants $\tau_1$ and $\tau_2$, respectively, estimated from fitting the data with bi-exponentially decaying function. Adapted from reference [55].

The typical z-scan experimental setup is shown in Fig. 15(b). A femtosecond laser beam is divided into two parts, one to excite the sample and the other for a reference to be used later for correcting the laser noise in the experimental data. A lens focuses the light onto the sample which is movable across the focal point along the optical axis so that for each z-position, the sample experiences a different incident laser intensity thereby varying the laser intensity from ~2 MW/cm$^2$ to 70 GW/cm$^2$. In Figs. 16(a) and 16(b), normalized transmittance for the $Au_{15}$-ITO and the $Au_{15}$-glass samples is presented. The signal from the OA z-scan shows optical limiting, i.e., reduction in transmission as the input beam intensity is increased, whereas the signal from CA z-scan shows a positive refractive nonlinearity (self-focusing effects), i.e., decrease in the transmitted intensity due to refraction as the sample approaches the focal point (z = 0) followed by an increase in intensity as it moves away from the focal point and towards the detector. We note from these results that the magnitude of change in transmission near the focal point is about 10 times larger for the $Au_{15}$-ITO sample as compared to the $Au_{15}$-glass sample. The continuous lines in Figs. 16(a) and 16(b) are theoretical fits to the data [44, 56] providing the two-photon absorption coefficient β of ~0.3 (3.0) cm/GW and the nonlinear refraction coefficient γ of ~2x10$^{-5}$ (6x10$^{-5}$) cm$^2$/GW for the $Au_{15}$-glass ($Au_{15}$-ITO) system. Clearly, the value of β (γ) for the $Au_{15}$-ITO sample is about 10 (3) times higher than that for the $Au_{15}$-glass sample. The z-scan experiments performed on bare ITO and glass plates under the same experimental conditions resulted in comparatively negligibly small values of the nonlinear coefficients.

The results for the transient differential transmission (ΔT/T) from pump-probe spectroscopy using pump-fluence of ~16 μJ/cm$^2$ and probe-fluence of ~1 μJ/cm$^2$ are presented in Fig. 16(c) and 16(d), on a linear-linear and log-linear scale to distinguish the regions of various dynamical time-constants. The following observations can be made from these results. Firstly, the polarity of the ΔT/T signals is negative which is in contrast to the positive signal (photo-bleaching) usually seen for larger colloidal gold nanoparticles at the surface plasmon resonance wavelength. In the present case, the negative signal corresponds to excited-state absorption and not the ground state bleaching [53]. The magnitude of the signal at zero probe-delay (t = 0) is a direct consequence of the magnitude of the excited state absorption at the probe wavelength of 790 nm. We can clearly see from Fig. 16(c) that the excited state absorption in the gold clusters is enhanced by a factor of ~8 for the clusters deposited on ITO as compared to those on the glass plate. This enhancement is similar to that observed from open aperture z-scan for the optical limiting performance of the $Au_{15}$-ITO system. The magnitude of ΔT(t=0)/T can be used to quantify the cascaded two-photon absorption process (one photon of 395 nm and the other of 790 nm) by estimating the corresponding nonlinear absorption coefficient $β_{(2)}$ via the relation $|\Delta T(t=0)/T| = β_{(2)}\sqrt{I_{pump} \times I_{probe}}$. The estimated value is $β_{(2)}$ ~ 0.05 cm/GW for $Au_{15}$-ITO and about 8 times smaller value for $Au_{15}$-glass.

The life-time of the photogenerated carriers in the excited state has been determined by fitting a bi-exponentially decaying function to the photo-induced absorption as shown in linear-linear and log-linear plots in Figs. 16 (c) and 16(d), respectively. It was found that the data for both the samples can be fitted nicely with an initial build-up time $τ_r$ ~ 200 fs, fast relaxation time-constant $τ_1$ ~ 700 fs and a slow relaxation time-constant $τ_2$ ~ 1 ns. The fast decay is attributed to the initial carrier relaxation from optically coupled states to the LUMO level followed by the slow decay via radiative and/or nonradiative processes from the LUMO level to the ground state. We find (data not shown) that the magnitude of the pump-induced absorption linearly increases with the excitation power (pump-fluence) indicating one-photon absorption processes at 3.15 eV. The carrier life-times remain pump-fluence independent. These results are similar to those reported for Au28 clusters [53].

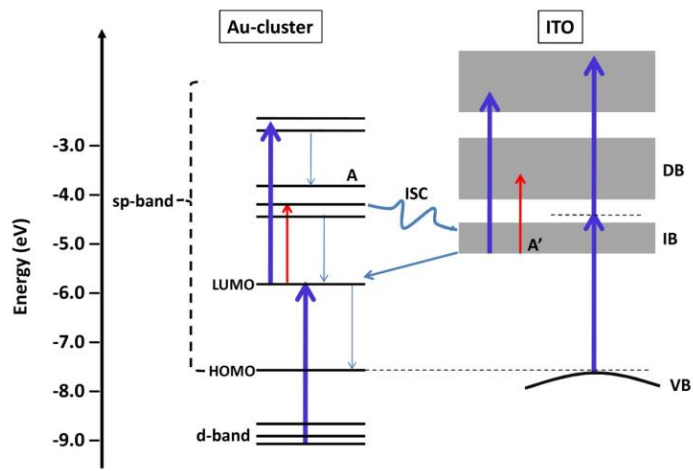

Fig. 17. Schematic of the electronic energy-diagram of Au$_{15}$ clusters coupled with the electronic energy-band diagram of ITO. The blue and red arrows represent the optical transitions by the pump and probe photon energies of ~3.15 and 1.57 eV, respectively. Possible excited-state intersystem coupling between Au$_{15}$ clusters and ITO is represented by the curved arrow and out of many possible interstate relaxation processes few are marked by thin downward arrows. Taken from reference [55].

The important outcome from this study was the observation of enhanced optical nonlinearity for the Au$_{15}$-ITO coupled system. A possible reason due to photoinduced energy transfer between the overlapped excited states of Au$_{15}$ and ITO has been outlined here. In the schematic energy level construction shown in Fig. 17, the molecular energy level diagram of the gold clusters [57] has been shown in the left side of the figure considering the absorption spectrum given in Fig. 15. The HOMO-LUMO gap of ~1.75 eV is very close to our probe photon energy of 1.57 eV. Three molecular like levels have been drawn at energies (wavelengths) of 2.1 eV (580 nm), 2.7 eV (458 nm) and 3.9 eV (318 nm). Similarly, on the right side in Fig. 17, we have drawn schematically, the energy bands in ITO as inferred from the indirect and direct band gap energies of ~2.4 eV (515 nm) and 3.6 eV (345 nm), respectively [58]. Since the enhancement in the excited state absorption (Fig. 16(c)) is similar to that in the optical limiting (Fig. 16(a)), hence it is natural to assume that the optical limiting in Au$_{15}$ clusters at 395 nm is also a two-step cascaded single photon absorption process. Whereas, a direct two-photon absorption in ITO at 3.15 eV (shown by two thick blue arrows in Fig. 17) from the top of the valance band (VB) to the continuum states bypassing the indirect band (IB) but via an intermediate virtual state is more probable. In the Au$_{15}$-ITO coupled system, due to close proximity of the energy levels of the molecular Au$_{15}$ clusters and energy bands of ITO, excited state electronic-coupling can occur. In that case the electrons created in the excited state of Au$_{15}$ (marked by A) can cross-over to the IB state of ITO which can later readily absorb single photons of either 3.15 or 1.57 eV photons as shown by vertical blue and red arrows in Fig. 17 leading to an enhanced combined two-photon absorption or photo-induced excited state absorption in the coupled system.

# 5 Direct observation of surface vibrations in nanoparticles

Plasmonic nanostructures' large absorption cross-sections, high sensitivity to geometry and refractive index changes, and the ability to localize electromagnetic fields into a subwavelength volume make them an ideal candidate for controlling nanoscale mechanical motion and sensing applications. By manipulating the geometry of these nanostructures, multiple vibrational states can be tailored and dynamically selected by acousto-plasmonic coherent control [59]. The shape-dependent confined vibrations of metal nanoparticles can be studied using various optical techniques, such as Raman spectroscopy, transient absorption spectroscopy and terahertz time-domain spectroscopy. In ultrafast time-resolved pump-probe spectroscopy, an incident femtosecond laser pulse initially excites the plasmons which quickly dephase in time and the hot electrons diffuse throughout the nanostructure at the Fermi velocity transferring their energy to the lattice. This results into rapid thermal expansion of the nanostructure generating coherent acoustic phonons by impulsive thermal excitation, [60, 61] which are imprinted as damped oscillations in the experimentally measured transient optical response signal.

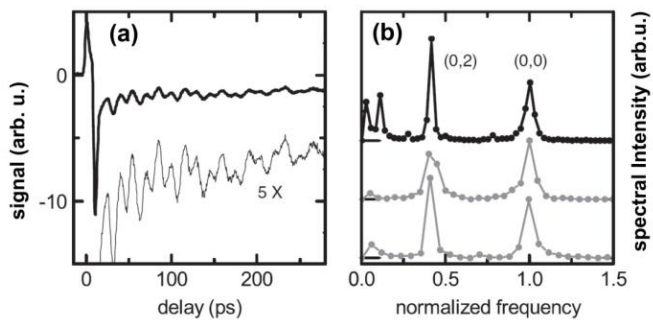

Fig. 18. (a) Delay scan of a single gold nanosphere with 45 nm diameter using interferometric pump-probe spectroscopy. (b) Power spectra of this particle's oscillations (top spectrum) and of two other particles of slightly different size. Frequencies and amplitudes are normalized to those of the (n, l) = (0, 0) mode (the absolute frequencies are 67, 59, and 63 GHz, from top to bottom). The low-frequency peak at ~28 GHz corresponds to the (n, l) = (0, 2) mode of the vibrations. Taken from reference [28] with permission.

The vibrational modes of a freely vibrating elastic sphere [62] are labelled by two integers, n, the harmonic order, i.e., the number of radial nodes, and l, the angular momentum number, which represents the angular dependence of the mode. As shown in Fig. 18, interferometric pump-probe measurements on single gold nanospheres using probe wavelength close to the surface plasmon resonance [28] revealed coherent excitation of a nonspherically symmetric mode of vibration (n, l) = (0, 2) along with the usual radial breathing mode of vibration (n, l) = (0, 0) that is observed very often for ensemble of particles where particle's environment is isotropic. The former can be generated by an isotropic heat pulse only if the spherical symmetry of the particle's expansion is broken either by the substrate or by the particle's shape.

For a rod-shaped nanoparticle, both the fundamental breathing mode of vibration along the short axis and the extensional mode of vibration along the long axis of the particle can be impulsively excited by laser-induced heating [63]. For gold nanorods in ensembles, these vibrational modes have been observed in many ultrafast pump-probe studies [63, 64], however, for a single nanorod, polarization of the pump-pulse can be used for selective excitation. Coherent acoustic vibrations of an optically trapped single gold nanorod as shown in Fig. 19 were reported by Ruijgrok et al., [30] which are very useful for determining the accurate elastic constants from the observed frequencies of the oscillations. The breathing mode involves a pure expansion and contraction (with change in volume) along the transverse radial directions of the rod, and depends on both the bulk and shear elastic moduli. Whereas the extensional mode is along the length of the rod involving dimensional changes in both transverse and longitudinal directions (without change in volume), and depends on the Young's modulus along the long axis of the rod.

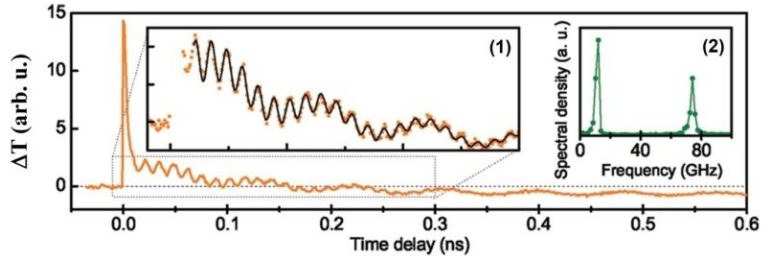

Fig. 19. Acoustic vibrations of a single gold nanorod. (a) Coherent oscillations in the transient transmission signal of a single 25 nm × 60 nm gold nanorod, optically trapped in water. The pump and the probe wavelengths were 785 nm and 610 nm, respectively (probe wavelength in the blue wing of the plasmon resonance). Inset (1): Zoomed-in of a part of the trace, fitted with a sum of two damped oscillations and slow cooling to extract the frequencies and damping times of the oscillations, inset (2): power spectral density of the oscillatory part of the vibrational trace. Taken from reference [30] with permission.

Both the extensional and breathing modes of gold nanorods having large aspect-ratio can be excited by femtosecond laser impulse heating in ultrafast time-resolved experiments. Experimentally, it is observed that for thinner rods (diameter < 20 nm), the breathing mode cannot be excited because the period of oscillations is comparable to the time-scale of laser heating [63]. Moreover, the detection of the extensional mode can be greatly improved by tuning the probe wavelength near the longitudinal surface plasmon resonance.

Our time-resolved pump-probe experiments performed upto probe-delays of ~400 ps on the colloidal suspensions of gold nanorods as discussed before, namely AuNR663, AuNR752, AuNR805 and AuNR854, revealed long time-period oscillations as shown in Fig. 20(a). These nanorods have diameter D of ~10 nm but growing length L so that the aspect ratios ($\zeta$ = L/D) are 2.8, 4.2, 4.7, and 5.2, respectively [39]. As can be seen from Fig. 20(b), the time-period of oscillations for the four gold nanorod samples is linearly dependent on the LSP wavelength. Since the LSP wavelength for these samples also varies linearly with the aspect ratio (Fig. 20(c)), it was confirmed that the observed coherent oscillations were due to extensional mode of vibrations (time-period $\tau_{ext}$) along the length of the nanorods. For the fundamental extensional mode, the relation between $\tau_{ext}$ and the Young's modulus is given [63] as $\tau_{ext} = 2L\sqrt{\rho/Y}$ where $\rho$ is the material density. Taking Y = 65 GPa and $\rho$ = 19.3 g/cm$^3$ for gold, [63] the average value of the nanorod length for the longest nanorods in our experiments, i.e., AuNR854 sample is estimated to be ~55 nm which is very close to that expected from D ~ 10 nm and $\zeta$ ~ 5.2.

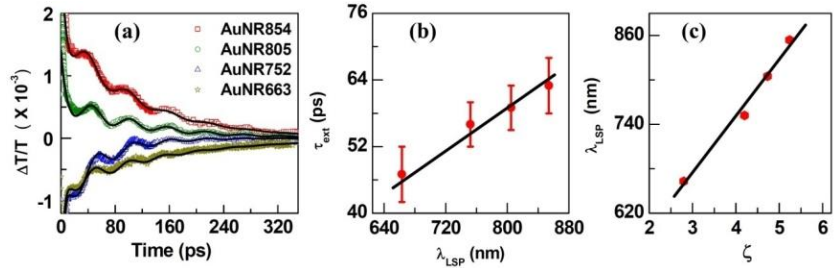

Fig. 20. Extensional mode of vibrations of gold nanorods observed using time-resolved pump-probe spectroscopy. (a) Coherent oscillations in the transient differential transmission spectra of colloidal suspensions of gold nanorods due to the extensional mode of vibrations, (b) correspondence between the time-period of the oscillations with LSP wavelength for the four samples, and (c) the linear dependence of the LSP wavelength on the nanorod aspect ratio. Adapted from reference [39] with permission.

The nanoparticle vibrational modes can also be characterized using time-domain terahertz (THz) spectroscopy as shown by Kumar et al., for silver nanoparticles [65]. Nearly spherical silver nanoparticles of mean diameter ~3.4 nm were prepared in a poly(vinyl alcohol) matrix; Figs. 21(a) and 21(b) show the transmission electron microscopy image and the absorption spectrum, respectively. Experimentally measured real and imaginary parts of the complex dielectric function ε are shown in Fig. 21(c) where the resonances due to the nanoparticle vibrations have been marked by arrows. A broad feature in the THz spectrum at frequency of ~1.1 THz is related to the crystalline lamellae in the thin polymer film.

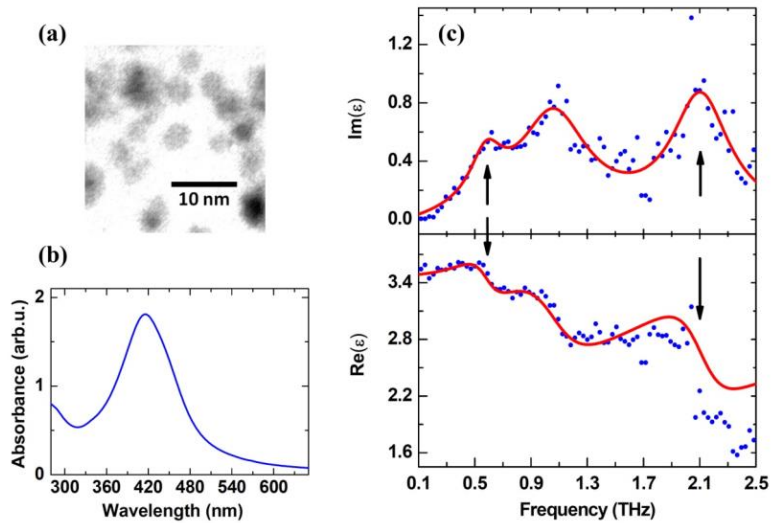

Fig. 21. Elastic vibrations of a silver nanospheres observed using terahertz time-domain spectroscopy. (a) Transmission electron micrograph of the silver nanoparticles dispersed inside thin film of poly(vinyl) alcohol, and (b) UV-Visible absorption spectrum of the nanoparticles. (c) Real and imaginary parts of the experimentally measured complex dielectric function ε in the THz frequency range showing vibrational resonances of the nanoparticles at frequencies marked by arrows. The resonance at ~1.1 THz is due to the crystalline lamellae in the polymer matrix. Adapted from reference [65] with permission.

The observed resonances at frequencies of ~0.6 and 2.12 THz are due to the spheroidal and toroidal vibrational modes of the nanospheres. These results clearly demonstrated for the first time that THz time-domain spectroscopy technique can be a complementary tool to the well-established Raman scattering spectroscopy for characterizing nanoparticles, particularly in the low-frequency regime where Raman spectroscopy is difficult to achieve the desired results due to strong Rayleigh scattering. We note that the phonons observed in the THz time-domain spectroscopy are the equilibrium phonons of the system unlike those created coherently by impulsive stimulated Raman scattering in the time-resolved pump-probe spectroscopy.

# 6      Summary and outlook

To summarize, ultrafast photoresponse of plasmonic nanostrucutres was described in this chapter by taking some of the interesting examples in the literature. The surface plasmon effect in metal nanostrucutres was discussed starting from the fundamental principles. We then showed that the ultrafast time-resolved spectroscopy of plasmonic nanostructures reveals the fundamental processes of electron scattering which help in releasing the excess energy of photoexcited electrons and the system returns to its equiillibrium. For gold nanorods, we showed that the ultrafast photo-bleaching and photoinduced absorption can be tuned and reversibly switched beween the two by varying the probe-wavelength across the LSP resonance and using the probe-fluence as the control parameter, respectively. A possible application of this novel tunability between the PB and PA could be an ultrafast optical switch where switching is controlled by the fluence of the second laser pulse. In other words, it shows potential in applications where ultrafast light controls the light itself. We also discussed the effect of surface plasmon resonance in enhancing the nonlinear optical properties of metal nanostructures. For a comparison, we presented a study on nonplasmonic gold nanoclusters where the enhanced optical limiting or photoinduced absorption property was discussed. It was found that about an order enhancement in the third order optical susceptibility is observed for the $Au_{15}$ clusters in contact with the ITO as compared to the clusters deposited on a $SiO_2$ glass plate. These results indicate a role of the excited state inter-system crossing between the electronic states of the clusters and the ITO to enhance the nonlinear response in the coupled system. Such coupled systems which are, of course, different from the extensively studied polymer-nanoparticle composites, need to be exploited much more in the future to have better understanding of the underlying physics of very small metal clusters in proximity of the tunable planar surface plasmon mode of a metal film or having an excited state overlap. Finally, we discussed time-reolved pump-probe spectroscopy and terahertz time-domain spectroscopy as efficient experimental tools for characterizing the confined acoustic phonons in the metal nanoparticles.

In recent years, there has been a growing interest to understand coherent vis-a-vis incoherent energy transfer between the surface plasmon polaritons in a metal and excitons in  molecule assemblies [66, 67]. The interplay between coherent and incoherent pathways leads to a significant modification of the radiative damping, opening up unexplored avenues in coherent active plasmonics to be studied by time-resolved spectroscopies. We expect that the studies related to the ultrafast energy transfer between molecular assemblies and surface plasmons in hybrid nanomaterials like two dimensional  graphene or transition metal dichalcogenides (e.g., MoS$_2$ layers) with metal nanostructures will attract increasing interest in coming years.

**Acknowledgements** We acknowledge financial assistance from Nanomission Project of Department of Science and Technology, Government of India.

# References

1. Krasavin, A. V.; Zheludev, N. I. Active Plasmonics: Controlling Signals in Au/Ga Waveguide Using Nanoscale Structural Transformations. Appl. Phys. Lett. 2004, 84(8), 1416-1418.
2. Renger, J.; Quidant, R.; Hulst, N. van; Palomba, S.; Novotny, L. Free-Space Excitation of Propagating Surface Plasmon Polaritons by Nonlinear Four-Wave Mixing. Phys. Rev. Lett. 2009, 103, 266802.
3. Rotenberg, N.; Betz, M.; Driel, H. M. van. Ultrafast All-Optical Coupling of Light to Surface Plasmon Polaritons on Plain Metal Surfaces. Phys. Rev. Lett. 2010, 105, 017402.
4. Zia, R.; Schuller, J. A.; Chandran, A.; Brongersma, M. L. Plasmonics: The Next Chip-Scale Technology. Materials Today 2006, 9, 20-27.
5. MacDonald, K. F.; Zheludev, N. I. Active Plasmonics: Current Status. Laser Photonics Rev. 2010, 4(4), 562–567.
6. Ru, E. C. L.; Blackie, E.; Meyer, M.; Etchegoin, P. G. Surface Enhanced Raman Scattering Enhancement Factors: A Comprehensive Study. J. Phys. Chem. C 2007, 111, 13794-13803.
7. Lal, S.; Link, S.; Halas, N. J. Nano-Optics from Sensing to Waveguiding. Nature Photonics 2007, 1, 641-648.
8. Khlebtsov, N. G.; Dykman, L. A. Optical Properties and Biomedical Applications of Plasmonic Nanoparticles. Journal of Quantitative Spectroscopy & Radiative Transfer 2010, 111, 1–35.
9. Clavero, C. Plasmon-Induced Hot-Electron Generation at Nanoparticle/Metal-Oxide Interfaces for Photovoltaic and Photocatalytic Devices. Nature Photonics 2014, 8, 95-103.
10. Maier, S. A. Plasmonics: Fundamentals and Applications. Springer (2007).
11. Garcia, M. A. Surface Plasmons in Metallic Nanoparticles: Fundamentals and Applications. J. Phys. D: Appl. Phys. 2011, 44, 283001-283020.
12. Bohren, C. F.; Huffman, D. R. Absorption and Scattering of Light by Small Particles. Wiley Interscience, 1998.
13. Mishchenko, M. I.; Travis, L. D.; Lacis, A. A. Scattering, Absorption, and Emission of Light by Small Particles. Cambridge University Press, New-York, 2004.
14. Burda, C.; Chen, X.; Narayanan, R.; El-Sayed, M. A. Chemistry and Properties of Nanocrystals of Different Shapes. Chem. Rev. 2005, 105, 1025-1102.
15. Noguez, C. Surface Plasmons on Metal Nanoparticles: The Influence of Shape and Physical Environment. J. Phys. Chem. C 2007, 111, 3806-3819.
16. Schmid, G. The Relevance of Shape and Size of Au$_{55}$ Clusters. Chem. Soc. Rev. 2008, 37, 1909-1930.
17. Qian, H.; Zhu, M.; Wu, Z.; Jin, R. Quantum Sized Gold Nanoclusters with Atomic Precision. Accounts of Chemical Research 2012, 45(9), 1470-1479.
18. Shibu E. S.; Pradeep, T. Quantum Clusters in Cavities: Trapped Au$_{15}$ in Cyclodextrins. Chem. Mater. 2011, 23(4), 989-999.
19. Mie, G. Articles on the Optical Characteristics of Turbid Tubes, Especially Colloidal Metal Solutions. Ann. D. Phys. 1908, 25, 377.
20. Gans, R. Uber Die Form Ultra Mikroskopischer Goldteilchen. Ann. Phys. 1912, 342, 881.
21. Johnson, P. B.; Christy, R. W. Optical Constants of the Noble Metals. Phys. Rev. B 1972, 6, 4370–4379.
22. Link, S.; El-Sayed, M. A. Spectral Properties and Relaxation Dynamics of Surface Plasmon Electronic Oscillations in Gold and Silver Nanodots and Nanorods. J. Phys. Chem. B 1999, 103, 8410-8426.
23. Novo, C.; Gomez, D.; Perez, -J. J.; Zhang, Z. Y.; Petrova, H.; Reismann, M.; Mulvaney, P.; Hartland, G. V. Contributions from Radiation Damping and Surface Scattering to the Linewidth of the Longitudinal Plasmon Band of Gold Nanorods: A Single Particle Study. Phys. Chem. Chem. Phys. 2006, 8, 3540-3546.

24. Stoll, T.; Maioli, P.; Crut, A.; Fatti, N. D.; Vallee, F. Advances in Femto-Nano-Optics: Ultrafast Nonlinearity of Metal Nanoparticles. Eur. Phy. J. B 2014, 87, 260-278.
25. Voisin, C.; Fatti, N. D.; Christofilos, D.; Vallee, F. Ultrafast Electron Dynamics and Optical Nonlinearities in Metal Nanoparticles. J. Phys. Chem. B 2001, 105, 2264-2280.
26. Hartland, G. V. Optical Studies of Dynamics in Noble Metal Nanostructures. Chem. Rev. 2011, 111(6), 3858-3887.
27. Kiel, M.; Mohwald, H.; Bargheer, M. Broadband Measurements of the Transient Optical Complex Dielectric Function of a Nanoparticle/Polymer Composite upon Ultrafast Excitation. Phys. Rev. B 2011, 84, 165121-165126.
28. Dijk, M. A. van; Lippitz, M.; Orrit, M. Detection of Acoustic Oscillations of Single Gold Nanospheres by Time-Resolved Interferometry. Phys. Rev. Lett. 2005, 95, 267406-267409.
29. Muskens, O. L.; Fatti, N. D.; Vallee, F. Femtosecond Response of a Single Metal Nanoparticle. Nano Lett. 2006, 6(3), 552-556.
30. Ruijgrok, P. V.; Zijlstra, P.; Tchebotareva, A. L.; Orrit, M. Damping of Acoustic Vibrations of Single Gold Nanoparticles Optically Trapped in Water. Nano Lett. 2012, 12(2), 1063-1069.
31. Baida, H.; Mongin, D.; Christofilos, D.; Bachelier, G.; Crut, A.; Maioli, P.; Fatti, N. D.; Vallee, F. Ultrafast Nonlinear Optical Response of a Single Gold Nanorod near Its Surface Plasmon Resonance. Phys. Rev. Lett. 2011, 107, 057402-.
32. Masia, F.; Langbein, W.; Borri, P. Polarization-Resolved Ultrafast Dynamics of the Complex Polarizability in Single Gold Nanoparticles. Phys. Chem. Chem. Phys. 2013, 15(12), 4226-4232.
33. Voisin, C.; Christofilos, D.; Loukakos, P.; Fatti, N. D.; Vallee, F.; Lerme, J.; Gaudry, M.; Cottancin, E.; Pellarin, M.; Broyer, M. Ultrafast Electron-Electron Scattering and Energy Exchanges in Noble-Metal Nanoparticles. Phys. Rev. B 2004, 69(19), 195416-195428.
34. Allen, P. B. Theory of Thermal Relaxation of Electrons in Metals. Phys. Rev. Lett. 1987, 59, 1460-1463.
35. Ashcroft, N. W.; Mermin, N. D. Solid State Physics. Holt, Rinehart and Winston, New York, 1976.
36. Cahill, D. G.; Ford, W. K.; Goodson, K. E.; Mahan, G. D.; Majumdar, A.; Maris, H. J.; Merlin, R.; Phillpot, S. R. Nanoscale Thermal Transport. J. Appl. Phys. 2003, 93(2), 793-818.
37. Fatti, N. D.; Vallee, F.; Flytzanis, C.; Hamanaka, Y.; Nakamura, A. Electron Dynamics and Surface Plasmon Resonance Nonlinearities in Metal Nanoparticles. Chem. Phys. 2000, 251, 215-226.
38. Anija, M.; Kumar, S.; Kamaraju, N.; Tiwari, N.; Kulkarni, S. K.; Sood, A. K. Ultrafast Dynamics of Gold Nanorods: Tuning Between Photo-Bleaching and Photo-Induced Absorption. Int. J. Nanosci. 2011, 10, 687-691.
39. Kumar, S.; Anija, M.; Sood, A. K. Tuning Ultrafast Photoresponse of Gold Nanorods. Plasmonics 2013, 8, 1477–1483.
40. Chemla, D. S.; Herritage, J. P.; Liao, P. F.; Isaacs, E. D. Enhanced Four-Wave Mixing from Silver Particles. Phys. Rev. B 1983, 27(8), 4553-4558.
41. Ricard, D.; Roussignol, P.; Flytzanis, C. Surface-Mediated Enhancement of Optical Phase Conjugation in Metal Colloids. Opt. Lett. 1985, 10(10), 511-513.
42. Bloemer, M. J.; Haus, J. W.; Ashley, P. R. Degenerate Four-Wave Mixing in Colloidal Gold as a Function of Particle Size. J. Opt. Soc. Am. B 1990, 7(5), 790-795.
43. Philip, R.; Kumar, G. R.; Sandhyarani, N.; Pradeep, T. Picosecond Optical Nonlinearity in Monolayer-Protected Gold, Silver and Gold-Silver Alloy Nanoclusters. Phys. Rev. B 2000, 62(19), 13160-13166.
44. Bahae, M. S.; Said, A. A.; Wei, T.-H.; Hagan, D. J.; and Stryland, E. W. V. Sensitive Measurement of Optical Nonlinearities Using a Single Beam. IEEE J. Quant. Electron. 1990, 26(4), 760-769.
45. Tokizaki, T.; Nakamura, A.; Kaneko, S.; Uchida, K.; Omi, S.; Tanji, H.; Asahara, Y. Subpicosecond Time Response of Third-Order Optical Nonlinearity of Small Copper Particles in Glass. Appl. Phys. Lett. 1994, 65(8), 941-943.
46. Luo, S.; Chen, Y.; Fan, G.; Sun, F.; Qu, S. Saturable Absorption and Reverse Saturable Absorption on Silver Particles with Different Shapes. Appl. Phys. A 2014, 117, 891-894.
47. Elim, H. K.; Yang, J.; Lee, J.-Y.; Mi, J.; Ji, W. Observation of Saturable and Reverse Saturable Absorption at Longitudinal Surface Plasmon Resonance in Gold Nanorods. Appl. Phys. Lett. 2006, 88, 083107.
48. Schaaff, T. G.; Knight, G.; Shafigullin, M. N.; Borkman, R. F.; Whetten, R. L. Isolation and Selected Properties of a 10.4 kDa Gold: Glutathione Cluster Compound. J. Phys. Chem. B 1998, 102(52), 10643-10646.
49. Schaaff T. G.; Whetten, R. L. Giant Gold−Glutathione Cluster Compounds: Intense Optical Activity in Metal-Based Transitions. J. Phys. Chem. B 2000, 104(12), 2630-2641.

50. Link, S.; Beeby, A.; FitzGerald, S.; El-Sayed, M. A.; Schaaff, T. G.; Whetten, R. L. Visible to Infrared Luminescence from a 28-Atom Gold Cluster. J. Phys. Chem. B 2002, 106(13), 3410-3415.
51. Jin, R. Quantum sized, Thiolate-Protected Gold Nanoclusters. Nanoscale 2010, 2(3), 343-362.
52. Yadav, B. D.; Kumar, V. Gd@Au15: A Magic Magnetic Gold Cluster for Cancer Therapy and Bioimaging. Appl. Phys. Lett. 2010, 97(13), 133701-1337-3.
53. Link, S.; El-Sayed, M. A.; Schaaff, T. G.; Wetten, R. L. Transition from Nanoparticle to Molecular Behavior: A Femtosecond Transient Absorption Study of a Size-Selected 28 Atom Gold Cluster. Chem. Phys. Lett. 2002, 356(3-4), 240-246.
54. Philip, R.; Chantharasupawong, P.; Qian, H.; Jin, R.; Thomas, J. Evolution of Nonlinear Optical Properties: From Gold Atomic Clusters to Plasmonic Nanocrystals. Nano Lett. 2012, 12, 4661-4667.
55. Kumar, S.; Shibu, E. S.; Pradeep, T.; Sood, A. K. Ultrafast Photoinduced Enhancement of Nonlinear Optical Response in 15-atom Gold Clusters on Indium Tin Oxide Conducting Film. Opt. Express 2013, 21, 8483-8492.
56. Kamaraju, N.; Kumar, S.; Sood, A. K.; Guha, S.; Krishnamurthy, S.; Rao, C. N. R. Large Nonlinear Absorption and Refraction Coefficients of Carbon Nanotubes Estimated from Femtosecond z-scan Measurements. Appl. Phys. Lett. 2007, 91(25), 251103-251105.
57. Zhu, M.; Aikens, C. M.; Hollander, F. J.; Schatz, G. C.; Jin, R. Correlating the Crystal Structure of a Thiol-Protected $Au_{25}$ Cluster and Optical Properties. J. Am. Chem. Soc. 2008, 130(18), 5883-5885.
58. Matino, F.; Persano, L.; Arima, V.; Pisignano, D.; Blyth, R. I. R.; Cingolani, R.; Rinaldi, R. Electronic Structure of Indium-tin-oxide Films Fabricated by Reactive Electron-Beam Deposition. Phys. Rev. B 2005, 72(8), 085437-085445.
59. O'Brien, K.; Kimura, N. D. L.; Rho, J.; Suchowski, H.; Yin, X.; Zhang, X. Ultrafast Acousto-Plasmonic Control and Sensing in Complex Nanostructures. Nature Communic. 2014, 5, 4042-4047.
60. Perner, M.; Gresillon, S.; März, J.; Plessen, G. von; Feldmann, J.; Porstendorfer, J.; Berg, K.-J.; Berg, G. Observation of hot-electron pressure in the vibration dynamics of metal nanoparticles. Phys. Rev. Lett. 2000, 85, 792–795.
61. Zijlstra, P.; Tchebotareva, A. L.; Chon, J. W. M.; Gu, M.; Orrit, M. Acoustic Oscillations and Elastic Moduli of Single Gold Nanorods. Nano Lett. 2008, 8, 3493–3497.
62. Nishiguchi, N.; Sakuma, T. Vibrational Spectrum and Specific Heat of Fine Particles. Solid State Communic. 1981, 38, 1073-1077.
63. Hu, M.; Wang, X.; Hartland, G. V.; Mulvaney, P.; Juste, J. P.; Sader, J. E. Vibrational Response of Nanorods to Ultrafast Laser Induced Heating: Theoretical and Experimental Analysis. J. Am. Chem. Soc. 2003, 125, 14925-14933.
64. Hartland, G. V.; Hu, M.; Wilson, O.; Mulvaney, P.; Sader, J. E. Coherent Excitation of Vibrational Modes in Gold Nanorods. J. Phys. Chem. B 2002, 106(4), 743-747.
65. Kumar, S.; Kamaraju, N.; Karthikeyan, B.; Tondusson, M.; Freysz, E.; Sood, A. K. Direct Observation of Low Frequency Confined Acoustic Phonons in Silver Nanoparticles: Terahertz Time Domain Spectroscopy. J. Chem. Phys. 2010, 133, 014502-014505.
66. Vasa, P.; Wang,W.; Pomraenke, R.; Lammers, M.; Maiuri, M.; Manzoni, C.; Cerullo, G.; Lienau, C. Real-time Observation of Ultrafast Rabi Oscillations Between Excitons and Plasmons in Metal Nanostructures with J-aggregates. Nature Photonics 2013, 7, 128-132.
67. Wang, W.; Vasa, P.; Pomraenke, R.; Vogelgesang, R.; Sio, A. D.; Sommer, E.; Maiuri, M.; Manzoni, C.; et al., Interplay between Strong Coupling and Radiative Damping of Excitons and Surface Plasmon Polaritons in Hybrid Nanostructures. ACS Nano 2014, 8, 1056-1064.